\documentclass[final,5p,times,twocolumn,authoryear]{elsarticle}
\pdfoutput=1

\usepackage{soul}
\usepackage{epsfig}
\usepackage[T1]{fontenc}
\usepackage{xcolor}
\usepackage[hyphens]{url}

\usepackage{booktabs} 
\usepackage{tablefootnote}

\DeclareRobustCommand{\VAN}[3]{#2}
\let\VANthebibliography\thebibliography
\def\thebibliography{\DeclareRobustCommand{\VAN}[3]{##3}\VANthebibliography}

\DeclareUnicodeCharacter{2009}{\,}

\usepackage{graphicx}	
\usepackage{amsmath}	
\usepackage{amssymb}	
\usepackage{newtxtext}
\usepackage{caption}
\usepackage{subcaption}
\usepackage{textcomp}
\usepackage{gensymb}
\usepackage{hyperref}
\usepackage{txfonts}
\usepackage{threeparttable}
\usepackage{lipsum}
\usepackage{soul}
\usepackage{float}
\usepackage{xcolor}

\begin{document}
\begin{frontmatter}
\title{STARFIRE: An algorithm for estimating radio frequency interference in orbits around Earth}


\author[add1,add2]{Sonia Ghosh}
\author[add1]{Mayuri Sathyanarayana Rao\corref{cor1}}
\ead{mayuris@rri.res.in}
\author[add1]{Saurabh Singh}
\cortext[cor1]{Corresponding author}

\address[add1]{Astronomy and Astrophysics, Raman Research Institute, C V Raman Avenue, Sadashivnagar, Bangalore, 560080, India}
\address[add2]{Kapteyn Astronomical Institute, University of Groningen, Landleven 12, 9747 AG, Groningen, The Netherlands}





\begin{abstract}
Ground-based  21-cm experiments targeting the global signal from the periods of Cosmic Dawn (CD) and Epoch of Reionization (EoR) are susceptible to adverse effects presented by i) the ionosphere ii) antenna chromaticity induced by objects in its vicinity  iii) terrestrial radio frequency interference (RFI). Terrestrial RFI is particularly challenging as the  FM radio band spanning over 88-108 MHz lies entirely within the frequency range of the CD/EoR experiments ($\sim 40-200$ MHz). Multiple space-based experiments have been proposed to operate in the radio-quiet zone on the lunar farside.  An intermediate option in cost and complexity is an experiment operating in space in an orbit around Earth, which readily alleviates the first two challenges. However, the effect of RFI in Earth's orbit on the detection of global signal needs to be quantitatively evaluated. We present STARFIRE -- Simulation of TerrestriAl Radio Frequency Interference in oRbits around Earth -- an algorithm that provides an expectation of FM seeded RFI at different altitudes over Earth. Using a limited set of publicly available FM transmitter databases, which can be extended by the user community, we demonstrate the use of the STARFIRE framework to generate a three-dimensional spatio-spectral cube of RFI as would be measured in Earth orbit. Applications of STARFIRE include identifying minimum RFI orbits around Earth, producing RFI spectra over a particular location, and generating RFI heatmaps at specific frequencies for a range of altitudes. STARFIRE can be easily adapted for different frequencies, altitudes, antenna properties, RFI databases, and combined with astrophysical sky-models. This can be used to  estimate the effect of RFI on the detection of global 21-cm signal from Earth-orbit, and hence for sensitivity estimates and experiment design of an Earth orbiting CD/EoR detection experiment. 
\end{abstract}

\end{frontmatter}



\section{Introduction}
\noindent The standard model of cosmology posits that the Universe has evolved through several important epochs. Of these, the Cosmic Dawn (CD)($30 \gtrsim z \gtrsim15$) and the following Epoch of Reionization (EoR)($15 \gtrsim z \gtrsim6$) are particularly important in terms of the evolution of baryonic matter. Over CD and EoR, intergalactic medium (IGM) transformed from being mostly neutral following the epoch of recombination, to being almost fully ionized by the end of EoR . The spin-flip, hyperfine transition of neutral hydrogen, with a rest wavelength of $\sim 21$-cm, can be used as a probe to study CD and EoR \citep{pritchard201221, fialkov2014observable}.

\noindent We focus here on the sky-averaged (or global) redshifted 21-cm signal which is quantified as the mean differential brightness temperature of the spin-temperature of neutral hydrogen and the background radiation temperature (of the cosmic microwave background -- CMB) and is given by the Equation \ref{eq:EoR}:

  \begin{equation}\label{eq:EoR}
    \begin{aligned}[b]
        & \delta {T_b}\simeq27x_{HI}(1+\delta_b)\left(\frac{\Omega_bh^2}{0.023}\right)\left[\left(\frac{0.15}{\Omega_mh^2}\right)\left(\frac{1+z}{10}\right)\right]^{1/2}\\
        & \left(\frac{T_{s}-T_{CMB}}{T_{s}}\right)\left[\frac{\delta_vv_{r}}{(1+z)H(z)}\right]^{-1} mK,
    \end{aligned}
  \end{equation}



\noindent wherein $z$ quantifies the cosmological redshift, $x_{HI}$ is the fraction of neutral gas, $T_s$ is the 21-cm spin temperature, $T_{CMB}$ is the CMB radiation temperature, $ \delta_b$ is the baryon overdensity, $H(z)$ is the Hubble parameter, $h$ is the Hubble constant in units of 100 km s$^{-1}$ Mpc$^{-1}$, $\Omega_b$ and $ \Omega_m $ are the critical baryon and matter density in the Universe respectively \citep{zaldarriaga200421}.  The spectral shape (turning points) and strength of the CD/EoR signal depends on the astrophysics of the early Universe in addition to cosmological parameters \citep{cohen2020emulating}. Thus, a precise measurement of the signal would enable an understanding of the processes involved in the formation of the first sources, and the subsequent re-ionization of baryonic matter in the Universe.

\begin{figure}
    \centering
    \includegraphics[width=9cm]{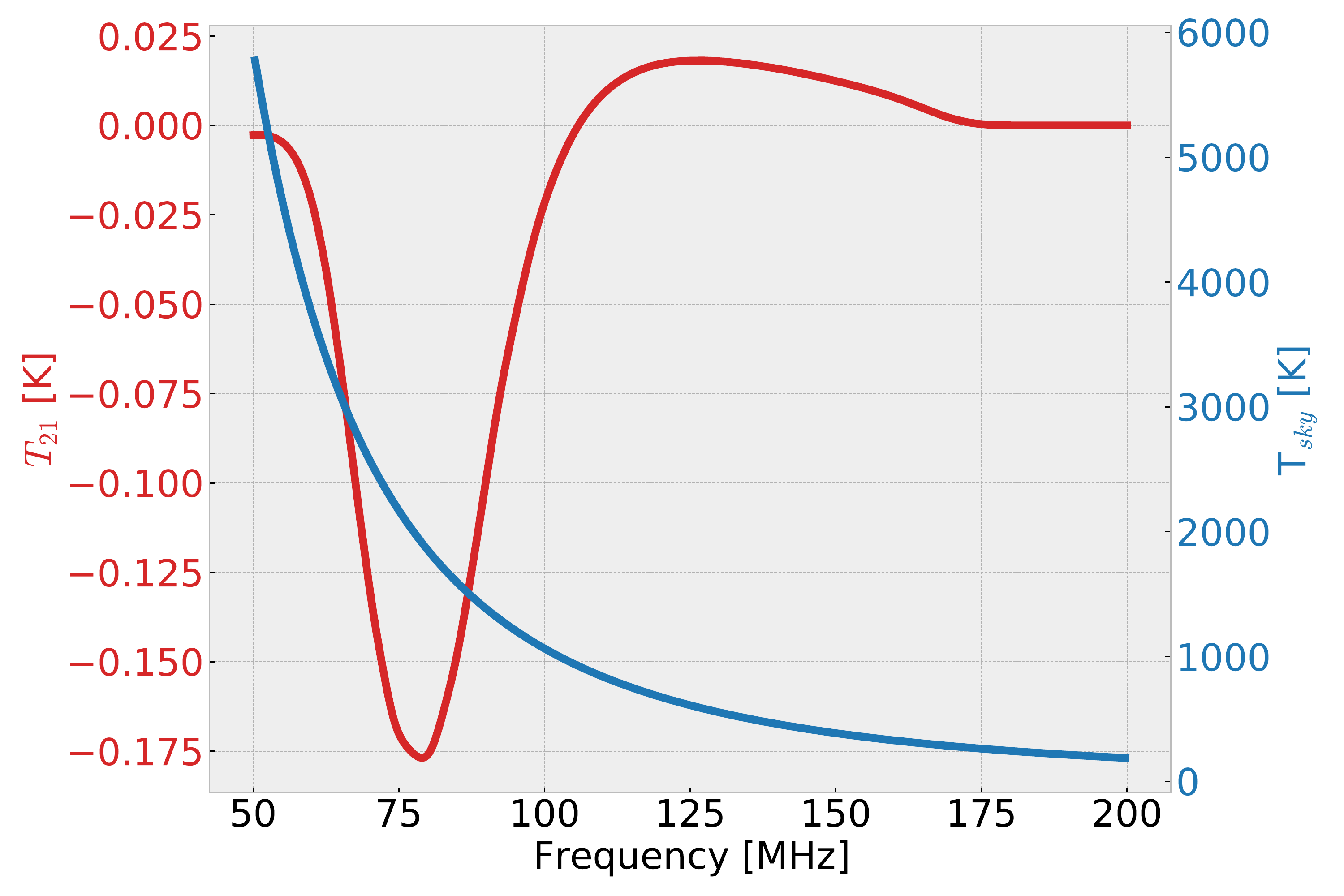}
    \caption{The red curve represents a fiducial model of 21-cm global signal from CD over $40-200$ MHz and the blue curve is the 24 hour averaged sky spectrum estimated using the GMOSS sky model \citep{rao2016gmoss}. The difference in spectral shape facilitates the separation of 21-cm signal from the Galactic foreground.}
    \label{fig:compare}
\end{figure}

\noindent From theory and existing observational constraints \citep{chen2004spin,fan2006observational} it is expected that the CD/EoR signal should be observable over the frequency range of $\sim 40 - 200$ MHz. Over this frequency range, the strength of the CD/EoR signal can be $4-6$ orders of magnitude weaker than foregrounds, dominated by Galactic synchrotron emission \citep{furlanetto2006cosmology}. However, the CD/EoR signal can be distinguished from the foreground owing to its predicted rich spectral features unlike the foreground component which is spectrally featureless or `smooth' \citep{rao2017modeling}. Figure \ref{fig:compare} displays the spectral signature of the CD/EoR signal ($\sim 10-100$ mK) as predicted from the "fiducial" model against the 24 hour sky-averaged radio sky spectrum estimated using GMOSS \citep{rao2016gmoss} having an  amplitude of $\sim 10^3K$, suggesting a dynamic range of $\sim 10^4- 10^6$.

Due to the global (sky-averaged) nature of the signal, it is expected that the CD/EoR signal can be detected using a single, well-calibrated radiometer with a wide beam. There are several ground-based experiments designed to target the global 21-cm signal from CD/EoR. These experiments include Shaped Antenna measurement of the background RAdio Spectrum (SARAS)  \citep{patra2013saras, singh2017first, subrahmanyan2021saras}, Experiment to Detect the Global Epoch of Reionization Signature (EDGES) \citep{bowman2018absorption}, the Radio Experiment for the Analysis of Cosmic Hydrogen (REACH) \citep{de2019reach}, Large Aperture Experiment to Detect the Dark Ages (LEDA) \citep{bernardi2015foreground}, Broadband Instrument for Global HydrOgen ReioNisation Signal (BIGHORNS) \citep{sokolowski2015bighorns} and Probing Radio Intensity at high-Z from Marion (PRIzM) \citep{philip2019probing}.
Notably, EDGES reported the detection of a deep and wide absorption trough centered at 78$\pm$1 MHz, with an amplitude of $500_{-200}^{+500}$mK \citep{bowman2018absorption}, raising possibilities of non-standard physical models if the detection was indeed cosmological in nature \citep{richard2018concerns, bradley2019ground, reis2021subtlety}. Measurements from SARAS-3, have refuted the astrophysical origins of the best-fit signal detected by EDGES with 95.3 \% confidence \citep{singh2022detection}. It is conceivable that the signal detected by EDGES could have origins in the instrument or residual systematics therein \citep{hills2018concerns, singh2019redshifted, sims2020testing}. The parameter space of CD/EoR, in particular the CD signal, remains wide open. A true high confidence detection would require measurements in a pristine environment using a precisely calibrated receiver with high level of control or knowledge of instrument systematics. 

Several efforts are underway to meet these stringent requirements, including improvements in ground-based experiments as well as proposals to operate experiments in the lunar farside owing to its radio-quiet environment. However, space based experiments, especially those that operate in the lunar farside come with additional challenges of design, operations, cost, and complexity.

An intermediate option is to operate a 21-cm CD/EoR experiment in space around Earth in orbits with low radio frequency interference (RFI). However, to explore the feasibility of such an option, RFI levels in Earth orbits need to be quantified. In lieu of a precise measurement of RFI over different locations and at different altitudes, an informed simulation can present a quantitative expectation of the same. In this context, we present STARFIRE -- Simulation of TerrestriAl Radio Frequency Interference in oRbits around Earth. 

The paper is organized as follows. Section \ref{sec:motivation} contains the motivation for STARFIRE. In Section \ref{sec:framework}, we outline the framework of STARFIRE. In Section \ref{sec:results}, we present  results from some simulations with STARFIRE. We discuss future directions that can be pursued in Section \ref{sec:discussion} and summarize in Section \ref{sec:summary}.

\section{Motivation}\label{sec:motivation}
\noindent Theoretically, the CD/EoR signal can be detected with over 90 \% confidence within 10 minutes of integration time using an ideal receiver \citep{rao2017modeling}. However, a high confidence detection remains elusive. This is because all ground-based experiments invariably face one or more of the three categorical challenges. These are 
\begin{enumerate}
    \item the frequency dependent and time varying effects of the Earth's ionosphere, which by means of refraction, absorption, and emission can potentially introduce chromatic effects that confuse signal detection \citep{vedantham2014, datta2016effects, shen2021quantifying}.
    \item `coupling' of antenna properties to the terrain over which it observes in addition to chromaticity intrinsic to broadband antennas \citep{anstey2022informing}. This also includes the boundary conditions of antenna edges \citep{bradley2019ground}, topographical effects of objects in the horizon \citep{bassett2021lost}, chromaticity introduced due to scattering off objects in the vicinity \citep{2022arXiv221203875S,2022RaSc...5707558R,2022JAI....1150001C}, or the inhomogeneous and stratified nature of the soil beneath the antenna reflector \citep{spinelli2022antenna}. 
    \item the ever increasing use of the radio spectrum for communication is leading to increased terrestrial radio frequency interference (RFI) occupancy in the frequency band of 40-200MHz. Based on the location of the observing site, terrestrial RFI may be detected directly via line of sight or reflected off objects, they can appear as narrow bright lines resulting in channel loss or as broadband contamination leading to an increased noise-floor level. This is dominated by RFI from FM broadcast stations, but may also include TV stations, airport and satellite communications, RFI scattering off objects beyond the horizon \citep{Offringa2013b}, among others. Broadly, the FM broadcast band occupies the frequency range of $88 - 108 $ MHz, with a few exceptions. This frequency band lies fully within the frequency range of experiments seeking to detect the CD/EoR signal. 
\end{enumerate}

The lunar farside is expected to provide a stable and pristine environment by significantly attenuating terrestrial RFI.  This has been observed decades ago at low radio frequencies by Radio Astronomy Explorer 2 (RAE-2)\citep{alexander1975scientific} and the WAVES instrument \citep{bougeret1995waves}. Recent electrodynamic simulations have also demonstrated that terrestrial RFI is suppressed by the moon in the lunar farside over the CD/EoR band with RFI signals attenuated by at least 80 dB \citep{bassett2020characterizing}  depending on the frequency and altitude of observation over the farside. Additionally, \cite{shi2022lunar} presents simulations of the global sky spectrum measurement and sensitivity of a global CD/EoR detection experiment predicted in lunar orbit.

In order to alleviate the detrimental impact of the ionosphere, the coupling effect of the antenna beam to its local environment and importantly terrestrial RFI faced by ground-based CD/EoR experiments, several experiments have been proposed to operate in the lunar farside either as landers or orbiters. Some of them are Dark Ages Reionization Explorer (DARE) \citep{burns2012probing}, Lunar Surface Electromagnetics Experiment (LuSEE) \citep{bale2020lunar}, Dark Ages Polarimeter PathfindER (DAPPER) \citep{burns2019dark}, Farside Array for Radio Science Investigations of the Dark ages and Exoplanets (FARSIDE) \citep{burns2019farside}, the Discovering the Sky at Longest wavelength (DSL) mission \citep{chen2021discovering}, among others. Of note is the Netherlands-China Low-Frequency Explorer (NCLE) \citep{2020AAS...23610203C} that has already made observations in lunar orbit, the published results of which are awaited. Joining the league of these lunar farside experiments is PRATUSH - Probing ReionizATion of the Universe using Signal from Hydrogen, a cosmology experiment from India that seeks to detect the CD/EoR signal in a lunar orbit. As a first step, PRATUSH in phase I is proposed to operate in Earth orbit to complement measurements from its ground based counterpart experiment SARAS \citep{singh2022detection} and to pave the way for the lunar orbiter experiment in its second phase. The baseline design of PRATUSH is optimized to operate over $55-110$ MHz, with a focus on detecting the poorly constrained signal from CD/EoR.
While the lunar farside is perhaps the most radio quiet region accessible to humankind, a precision experiment in such a location is associated with high cost and complexity compared to operating in space in Earth orbit. A low frequency cosmological experiment operating in free-space over Earth certainly alleviates two of the three challenges faced by ground-based experiments (namely those of ionosphere induced chromaticity and coupling effects arising from antenna-environment interactions), but terrestrial RFI remains an inevitable contaminant even in such a case. This necessitates a quantitative assessment of the RFI environment in space around Earth. The RFI levels (amplitude and spatial-frequency occupancy) can be used for an informed design of a space-based CD/EoR signal detection experiment orbiting Earth. Based on the RFI levels estimated, such an experiment can make a high confidence signal detection, or improve data quality compared to ground-based experiments. In the worst case scenario wherein the CD/EoR signal detection is limited entirely by RFI as measured by ground-based and Earth orbiting space-based experiments, early Earth orbiters can demonstrably justify the necessity to observe in more radio quiet locations in space, such as the lunar farside. Furthermore, an Earth-orbiting experiment can serve as a technology and sensitivity demonstrator to a future planned lunar orbiter mission and serve to mitigate risks early. 
Intuition might suggest that by being at large distances from terrestrial RFI sources, the RFI signal strength decreases thus favoring CD/EoR detection compared to ground-based experiments. However, the greater distance also implies a larger field-of-view (FOV) and thus increases the vulnerabilty of the experiment to a larger number of RFI sources. The trade-off between the free space path loss and FOV needs to be quantified to ask the question - "Are there any favorable orbits around Earth that are conducive to CD/EoR signal detection?".

The above question motivates evaluating the expected levels of RFI in Earth-orbit for radio frequency cosmology experiments like PRATUSH. In particular, we focus on interference in the FM band as this is among the worst-offenders in terms of terrestrial RFI in the CD/EoR band. We present an algorithm - STARFIRE, which provides a framework to estimate RFI levels in the FM band at different altitudes and locations over Earth. Though designed in the context of CD/EoR signal detection, STARFIRE can be used for any application that requires an understanding of the RFI landscape in space over Earth, as desired by the end-user, and maybe readily extended to other frequencies based on the availability of input datasets. 

\section{STARFIRE Framework} \label{sec:framework}

\begin{figure*}
    \centering
    
    \includegraphics[width=22cm,angle=90]{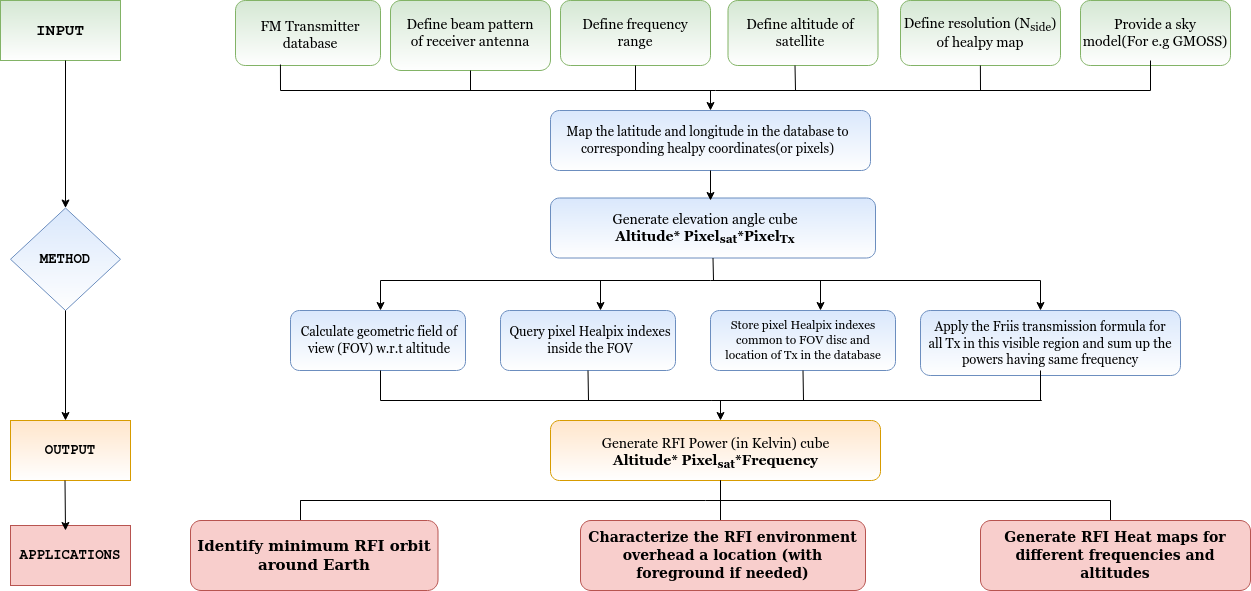}
    \caption{Schematic representation of the architecture used in this work to generate the output data cube of RFI power having dimensions $N_{altitude} \times N_{pixel} \times N_{frequency}$. }
      \label{fig:flowchart}
    
\end{figure*}

\noindent The user defined inputs to STARFIRE are a database of FM transmitting stations, altitude of the receiving satellite, beam-pattern of the receiving and transmitting antenna, frequency range and resolution, pixel resolution of the Earth map, and a sky-model (optional). The output of STARFIRE is a spectral cube containing RFI power with dimension $N_{altitude} \times N_{pixel} \times N_{frequency}$. The architecture and steps followed by the STARFIRE algorithm are shown in Figure \ref{fig:flowchart}. The output spectral cube can be sliced along appropriate axes to provide useful metrics of RFI (FM) occupancy, including all-sky RFI heatmaps at specific frequencies and altitudes, variation in total RFI power with height over a specific pixel, or an RFI frequency-spectrum over a particular location at a specific altitude. The algorithm can also simulate the total sky-spectrum including the galactic and extra-galactic emission (foreground), RFI, and the global CD/EoR signal with appropriate inputs. We use PRATUSH as the reference experiment when assuming instrument parameters where appropriate. STARFIRE can be easily adapted to other

\begin{figure}[H]
    \centering
    
    \includegraphics[width=7cm]{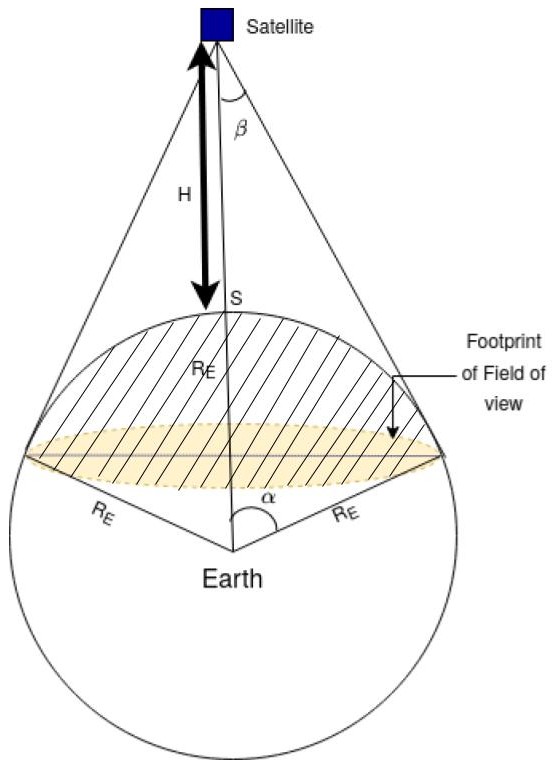}
    \caption{Geometry for the calculation of the FOV of the satellite at an altitude $H$. $R_{E}$ is the mean radius of the Earth = 6371 km. The satellite FOV $(2\beta)$ angle and the Earth central angle $(\alpha)$ are represented. }
     \label{fig:FOV}
    
\end{figure}

\noindent experiments and applications of the user's choosing. Customising the user inputs to STARFIRE can guide experiment design, including choice of orbit, designing an antenna with a beam pattern to minimize the RFI, to name a few. The ultimate objective of STARFIRE is to study the impact of RFI in FM band from the Earth's orbit on PRATUSH-like experiments to find optimal Earth-orbits, if any, for a low-cost (compared to lunar orbit) Earth orbiting CD/EoR detection experiment or precursor to a future lunar experiment . 
\subsection{STARFIRE GEOMETRY}\label{sec:notation}

Although there can be several sources of RFI, we focus on RFI in the FM radio band. We denote FM transmitting stations on Earth as transmitters and the element (antenna, experiment) in space which intercepts the power from this radiation as the receiver or receiving satellite. We assume that the Earth is a perfect sphere with radius $R_{E}$. The line joining the satellite and the center of the Earth passing through the surface of the Earth at a point S is called the subsatellite point. Figure \ref{fig:FOV} shows the geometry for the calculation of the FOV of the satellite from a particular altitude and its geometric footprint. We adopt a HEALPix \citep{Zonca2019} pixelization scheme for Earth. We clarify here that the HEALPix index for every subsatellite point is the same as the corresponding pixel in space, overhead. Each receiver location or pixel in space is denoted with the same pixel number as that of the subsatellite point $S$. The half-angle subtended between the receiver satellite and tangents to Earth is the nadir angle denoted by $ \beta$ in radians:
  

\noindent
 \begin{flalign}
     &\mathrm{\beta= sin^{-1}\left(\dfrac{R_E}{R_E+H}\right)},&
 \end{flalign}

\noindent where $R_E$ is the mean radius of the Earth = 6371 km and $H$ is the altitude of the satellite above Earth at the subsatellite point $S$ in km. Not considering diffraction, a receiving satellite can at most receive RFI from all transmitters on the Earth that lie in its geometric FOV. A directive antenna on the receiving satellite, on the other hand, can only receive RFI from a subset of the transmitters in the FOV, depending on the receiving antenna beam pattern, in addition to the FOV.  The FOV is strictly a geometrical quantity, determined by the height of the satellite from the Earth's surface and the Earth's mean radius. In the limiting case, a receiving satellite can at most see RFI from transmitters in one hemisphere. From  trigonometry, we find that FOV=2$\beta$ and the area projected on Earth by the FOV has a circular cross  diameter 
\noindent
\begin{flalign} \label{eq:beta}
&\mathrm{Diameter\hspace{1mm} of\hspace{1mm} FOV = 2 \alpha R_E }.&
\end{flalign}

\noindent which depends on the central angle $ \alpha$ (in radians) given by Equation \ref{eq:alpha}:

\begin{flalign}\label{eq:alpha}
&\mathrm{\alpha = cos^{-1}\left(\dfrac{R_E}{R_E+H}\right)}.&
\end{flalign}



\subsection{IMPLEMENTATION}\label{sec:implementation}

The inputs to STARFIRE are namely :
\begin{itemize}

    \item The altitudes over which the RFI expectation is to be computed. The defaults are 400 km, 3795 km and, 36000 km, the two extremes representing nominal values for altitudes of Low Earth Orbit and Geo Stationary satellites, respectively.
    \item The frequency range and resolution over which the computation is to be performed, equivalently the frequency bins of the output spectrum required. The default frequency range is $55 - 110$ MHz with a resolution of $\sim 244$ kHz, according to the specifications of PRATUSH.
    \item FM transmitter database. The default option of a limited FM transmitter database is described in Section \ref{sec:database}.
    \item The number of pixels or spatial resolution of the map defined by the HEALPix term $N_{side}$. The default value of  $N_{side}$ is 16 for both the sky and Earth. This corresponds to 3072 pixels of resolution $\sim 3.66\degree$ each. While the spatial resolution can be increased or decreased, as applied to global 21-cm detection experiments with typically large (well over $\sim 60\degree$) beams, a resolution of $N_{side}$= 16 is a reasonable choice. Higher spatial resolution with the limited database presented herein will likely result in some changes in results for simulations at lower altitudes and at boundary pixels, and would not significantly alter the results for large altitudes. However, we acknowledge that higher resolution is useful for generic applications. STARFIRE-2 will include a statistical model of the transmitter database and will present the option to work with finer resolution in a more meaningful way.
\end{itemize}
In addition to the above, optional inputs to STARFIRE are
\begin{itemize}
    \item The beam pattern of the transmitting antennas. The default beam pattern of the transmitter is assumed to be isotropic and frequency independent. It is usually difficult to obtain information about the radiation pattern of antennas used in FM transmitters from the publicly available databases. Thus, by default, all the transmitters are assumed to radiate isotropically in the default implementation. The algorithm has the flexibility to incorporate preferred radiation patterns for the transmitters.
    \item The beam pattern of the receiving antenna. The default beam pattern of the receiver is assumed to be $cos^2\theta$ and frequency independent, where $\theta$ denotes the elevation angle defined with respect to the satellite azimuthal plane.
    \item A model of the low-frequency radio sky such as Global Model for the Radio Sky Spectrum (GMOSS) \citep{rao2016gmoss}, the Global Sky Model (GSM) \citep{2010ascl.soft11010D} or the improved GSM \citep{zheng2017improved}.
\end{itemize}
 The output of STARFIRE is a three-dimensional data cube that can be utilized to generate RFI spectra, RFI heatmaps, and identify optimal orbits orbits around Earth that maximize the Figure of Merit (FOM) as described in Section \ref{sec:FOM}.

 \begin{table}
 
	\centering
\begin{threeparttable}
	\caption[\TeX{} engine features]{%
	The table shows the countries and the number of FM transmitters in the database.}
	\label{tab:data}
	\begin{tabular}{l|r} 
		\hline
		Country & Number of FM Transmitters \\
		\hline
		Australia & 2664 \\
		Canada & 8443 \\
		Germany & 2500 \\
		Japan(Tokyo) & 21 \\
		South Africa & 1791(1000)\tnote{1}\\
		USA & 28072\\
		\hline
	\end{tabular}
\begin{tablenotes}
   \item[1] \footnotesize Due to limitations in the performace of optical character recognition based  extraction on this file we have used a manually entered subset of 1000 values from the South Africa database for this work. 
  \end{tablenotes}
\end{threeparttable}
  
\end{table}

\subsection{DATABASE} \label{sec:database}
For the present implementation of STARFIRE, we use a  sample FM transmitter database from around the globe. The database contains information about the location (latitude, longitude) of the transmitters in degrees, effective radiated power (ERP) in Watt, and frequency of transmission in MHz. In addition to the availability of the data, the current model ensures that the database includes geographic regions that are spread across the globe to include a wide sample of RFI spatial-occupancy as seen by a satellite over Earth. For our analysis, we use publicly accessible databases from the countries given in Table \ref{tab:data}. We acknowledge the sources from which we have obtained the data and refer the reader to the websites for the respective countries in Section \ref{sec:data}. We claim no responsibility for the veracity of the data presented in these sources. The user may extend or change the input database (including location, frequency, and power) as per their requirement and computational resources available. We acknowledge that the results of the algorithm is highly dependent on FM transmitters used as the input to the algorithm and that obtaining an accurate and complete database of transmitters from around the world is extremely challenging. We also recognize that terrestrial RFI in the CD/EoR band is
\begin{figure}[H]
    \centering
    \includegraphics[width=6cm]{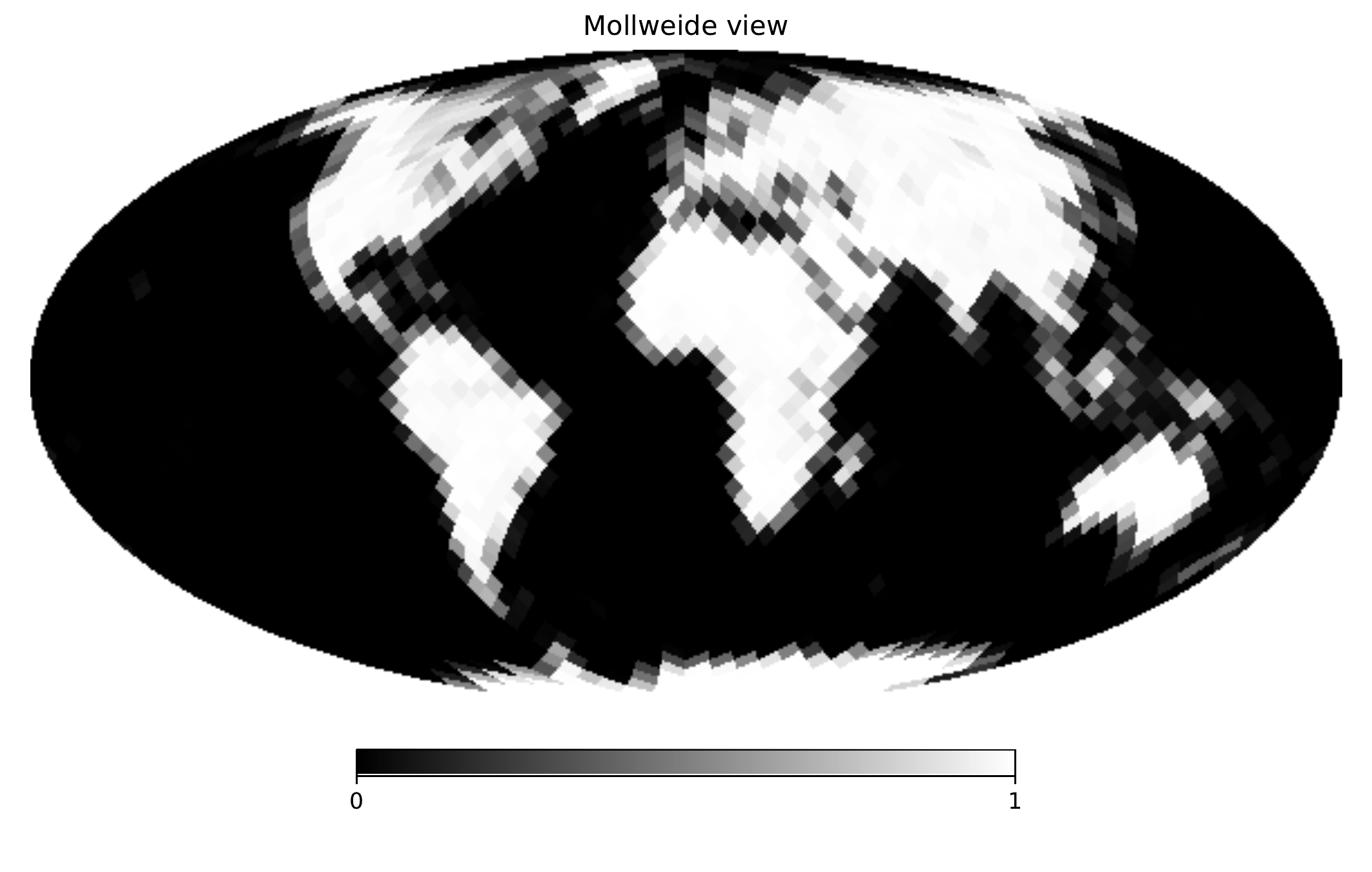}
    \caption{Outline map of World continents in Mollweide projection with $N_{side}$=16. The color bar represents the fractional land mass per pixel, 0 being water only and 1 being land only.}
     \label{fig:continent_layout}
\end{figure}
\begin{figure}[H]
     \centering
     \begin{subfigure}[b]{0.23\textwidth}
         \centering
         \includegraphics[width=\textwidth]{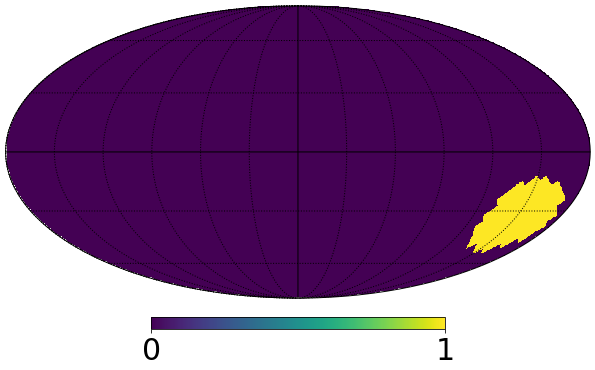}
         \caption{Altitude = 400 km}
         
     \end{subfigure}
     \hfill
     \begin{subfigure}[b]{0.23\textwidth}
         \centering
         \includegraphics[width=\textwidth]{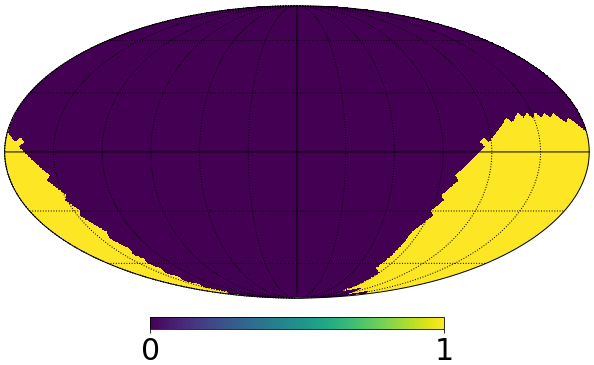}
         \caption{Altitude = 3795 km}
     \end{subfigure}
     \hfill
     \begin{subfigure}[b]{0.23\textwidth}
         \centering
         \includegraphics[width=\textwidth]{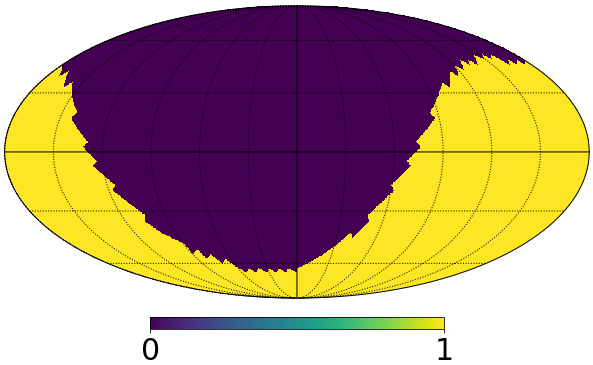}
         \caption{Altitude = 36000 km}
     \end{subfigure}
     
        \caption{ Geometric footprint that defines the FOV of a satellite at three different altitudes overhead Sydney (33.8688\degree S, 151.2093\degree E). The Earth is represented in Mollweide projection. This geometric FOV effectively acts as a masking function, wherein pixels in yellow (weighting of 1) are visible to the satellite, and those in deep purple (weighting of 0) are not.}
        \label{fig:geometric}
       
\end{figure}
\noindent not limited solely to FM transmitters, and can include other sources of RFI such as television (TV), transformers, satellite and aircraft communication. The utility of the STARFIRE algorithm presented herein lies in estimating the effect of any input database of RFI from stationary sources (earth-centered inertial frame) on spectra as observed in space around Earth, accounting for free-space loss and receiver parameters including position and antenna beam pattern. While users can build the input database by collecting this information, an alternative option is to build a statistical model of the distribution of the FM transmitters as will be presented in a follow up paper on STARFIRE-2. STARFIRE-2 will present a statistical model to populate the FM transmitter database. It will populate positions, frequencies and transmitted powers of FM transmitters using underlying correlations between entries in the ``true” database and proxies of FM connectivity such as population density or global connectivity index. As the true FM transmitter database from across the globe improves with inputs from the user community, the statistical model will continue to improve with reasonable tests on assumptions of the underlying correlations as proxies. It is expected that this will provide a more complete and realistic prediction of RFI in space using STARFIRE to suit the user's application. For instance we expect that the heatmaps will be more ``filled” for all altitudes with fewer clear blue patches. However, one might reasonably assume that even in the statistically realized model of the global transmitter database the oceans would look largely clear of RFI at sufficiently low altitudes. A further discussion of the statistical model is beyond the scope of this work. 
      
\begin{figure}
    \centering
    
    \includegraphics[width=7cm]{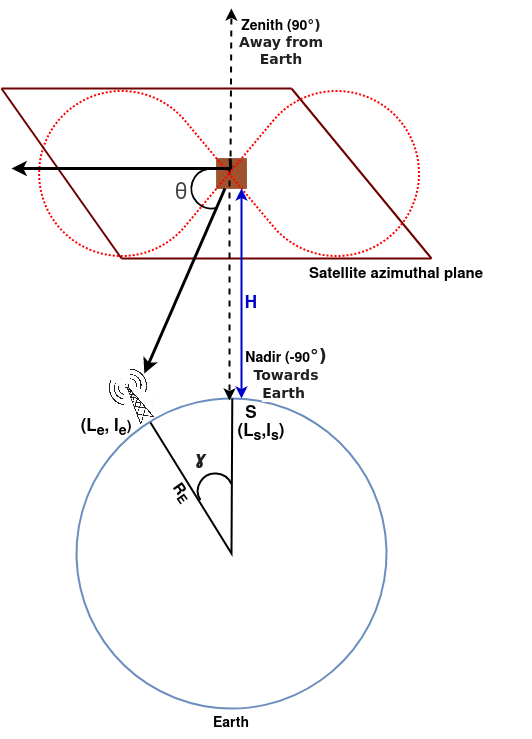}
    \caption{Schematic (not-to-scale) of the elevation angle $\theta$ of the satellite (receiver) antenna beam used in calculation. The beam of the receiving antenna is assumed to have a $cos^2\theta$ radiation pattern measuring downward from the azimuthal plane defined by the satellite.}
     \label{fig:el}
    
\end{figure}

 \noindent For quick reference and easy visualization of the subsequent Healpy maps in Mollweide projection, a geographical continent map is presented in Figure \ref{fig:continent_layout}. The grayscale image indicates fractional landmass per pixel corresponding to $N_{side} = 16$.
 Figure \ref{fig:geometric} shows the geometric footprint or FOV of such a receiving satellite for altitudes of 400 km, 3795 km and 36,000 km, overhead Sydney (33.8688\degree S, 151.2093\degree E). The power (in Kelvin) at each pixel is the isotropic beam averaged value received when overhead the corresponding pixel. 
\noindent We specify the radiation pattern of the receiving antenna in a coordinate system (as illustrated in Figure \ref{fig:el}) that is defined with respect to the subsatellite point \citep{pratt2019satellite}. The elevation angle ($\theta$) is the angle between the satellite's azimuthal (horizontal reference) plane and the transmitting station on Earth.  As per the convention adopted in STARFIRE, when the satellite is overhead a subsatellite point, the elevation angle of the satellite's antenna beam is $-90\degree$. Correspondingly the elevation towards the sky-ward zenith angle is $90\degree$. Equation \ref{eq:theta} provides the formal definition for elevation angle:

\begin{flalign}\label{eq:theta}
&\theta =\mathrm{tan^{-1}\left(\dfrac{\left(\dfrac{R_E+H}{R_E}\right) - cos \gamma}{sin \gamma}\right)},&
\end{flalign}

\noindent wherein, 
\begin{flalign}
&\mathrm{cos} \hspace{1mm} \gamma= \mathrm{cos(L_s)}\hspace{0.8mm}\mathrm{cos( L_e )}\hspace{0.8mm}\mathrm{cos(l_{e} - l_{s} )} + \mathrm{sin(L_s )\hspace{0.8mm} sin(L_e )},&
\end{flalign}

\begin{figure}
     \centering
     \begin{subfigure}[b]{0.23\textwidth}
         \centering
         \includegraphics[width=\textwidth]{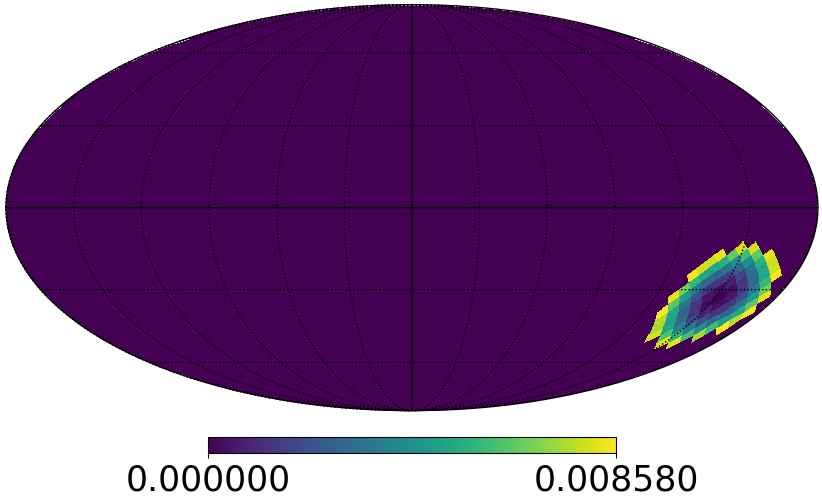}
         \caption{Altitude = 400 km}
         
     \end{subfigure}
     \hfill
     \begin{subfigure}[b]{0.23\textwidth}
         \centering
         \includegraphics[width=\textwidth]{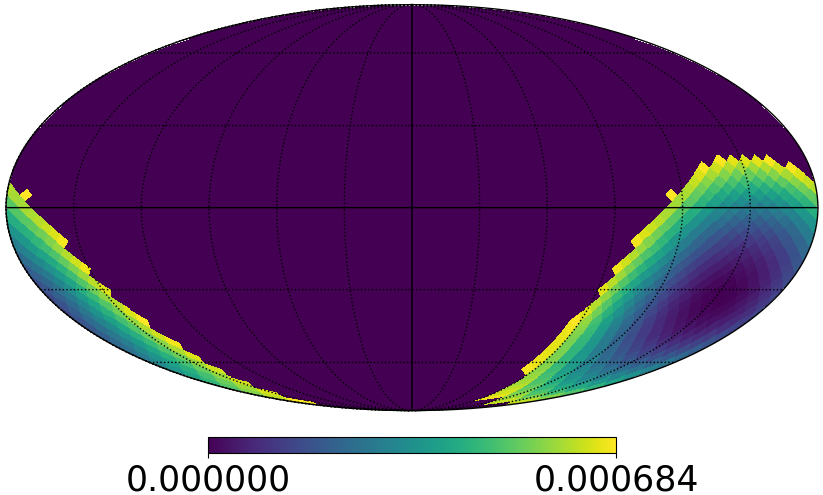}
         \caption{Altitude = 3795 km}
     \end{subfigure}
     \hfill
     \begin{subfigure}[b]{0.23\textwidth}
         \centering
         \includegraphics[width=\textwidth]{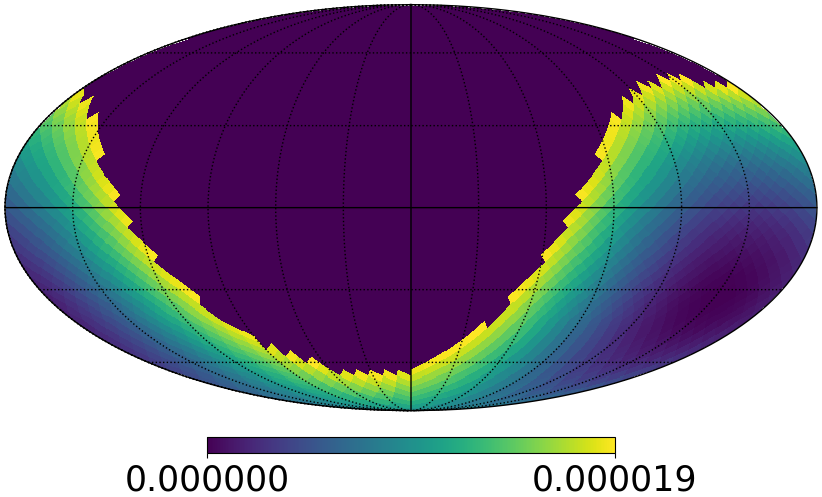}
         \caption{Altitude = 36000 km}
          \end{subfigure}
        \caption{The footprints of a satellite at three different altitudes (with the null towards the subsatellite point) weighted by the dipole beam pattern of the receiving antenna. The maps are in Mollweide projection wherein the satellite is assumed to be overhead Sydney (33.8688\degree S, 151.2093\degree E) .}
        \label{fig:beamweight}
       
\end{figure}

\begin{figure}
    \centering
    \includegraphics[width=8cm]{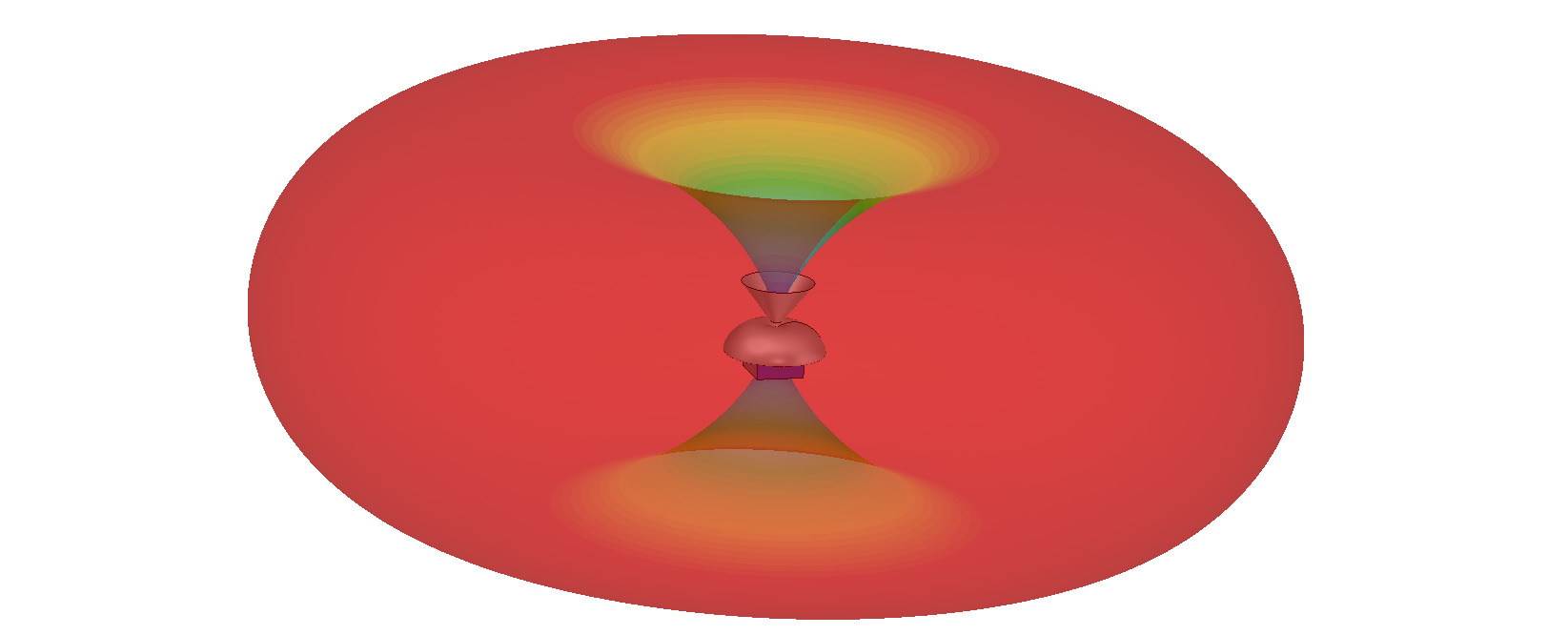}
    \caption{The 3-dimensional far-field radiation pattern of PRATUSH at a nominal frequency.}
    \label{fig:PRATUSH_field}
\end{figure}

\noindent where $ \gamma$ is the angle between the transmitter location on Earth and the satellite, $ L_s$ and $L_e$ is the geographical latitude in degrees that the subsatellite point and the transmitter is north from the equator respectively, $l_s$ and $l_e$ is the geographical longitude in degrees that the subsatellite point and the transmitter is west from the Greenwich meridian respectively. In Figure \ref{fig:beamweight}, the receiver antenna is assumed to have a $cos^2\theta$ beam pattern. The choice of the beam pattern is governed by the fact that the PRATUSH beam closely resembles a $cos^2\theta$ pattern in elevation, with nulls towards the zenith (sky) and nadir (Earth). PRATUSH uses an electrically small monopole antenna over a finite, profiled ground plane. Although this results in a wide beam with significant sensitivity towards the Earth (or the Moon, when in lunar orbit) in addition to the sky, this design is optimized to reduce beam chromaticity and minimize inflections in antenna return loss. The PRATUSH antenna beam is shown in Figure \ref{fig:PRATUSH_field} at a nominal frequency in the 55-110 MHz band, and has a variation at the level of a few percent over the band. For instance, when the satellite is overhead a location (in this example Sydney), the elevation angle towards that location is -90\degree  and the corresponding field strength is zero. Similarly the elevation angle towards the edge of FOV for a satellite at an altitude of 400 km is  $\sim-84.23$\degree  and hence the dimensionless beam weight is $\sim$ 0.009 (see Figure \ref{fig:beamweight}a ).

Transmitters (FM stations) and receivers (satellite) are assigned pixel numbers (adopting the HEALPix formalism) corresponding to their respective locations. Next, we calculate the geometric FOV for each position of the receiving satellite corresponding to the range of altitudes for which the algorithm is configured. From this we extract the pixel numbers of FOV, that is, we extract a set of pixels numbers of the locations on Earth that lie in the geometric footprint of the satellite. We proceed to extract the intersection of the sets of pixel numbers in the transmitter (FM) database and the receiver FOV as estimated in the previous step. The Friis transmission formula (Equation \ref{eq:FT}), is then used to estimate the power intercepted by the receiver from the transmitters in this intersection set:

\begin{flalign}\label{eq:FT}
&\mathrm{P_r(\nu)=\dfrac{P_t(\nu)G_r(\nu)G_t(\nu)c^2}{\nu^2(4\pi R)^2}},&
\end{flalign}
where $\nu$ is the frequency of interest, $ P_r$ is power at the receiving antenna in Watts, $ P_t$ is the output power of transmitting antenna in Watts, $ G_r$ is gain of the receiving antenna, $G_t$ is the gain of the transmitting antenna, $R$ is the mean geometric distance between the transmitting antenna and receiving antenna in meters. While nominal heights of Low Earth Orbits (such as 400 kilometre or above) still lie within the ionosphere, these lie above the D-layer and at least in the upper reaches of the F-layer, if not beyond. While the presence of the ionosphere can induce chromatic effects pertaining to cosmological and astrophysical signals \citep{10.1093/mnras/stt1878}, at heights typical of satellites these effects should be considerably lesser. The ionosphere can also refract, absorb, and attenuate the FM signals as seen by the satellite, however we relegate this to a second order problem and focus on the RFI power received directly as seen by the satellite without ionospheric screening. The received RFI power in Watts is converted to the equivalent temperature $T_{bright}$ \citep{Nyquist_noise}. For a given bandwidth $\Delta\nu$ determined by the user defined frequency resolution , the equivalent brightness temperature $T_{bright}$ (in Kelvins) for the power received by an antenna is given by:
\begin{flalign}\label{eq:kelvin}
&\mathrm{T_{bright}=\dfrac{P_r}{k\Delta\nu}},&
\end{flalign}

\noindent where k denote the Boltzmann constant.
For each receiving satellite location, the brightness temperature of the corresponding beam weighted RFI power from all transmitters in the FOV (for each altitude) is frequency binned based on the user defined frequency range and resolution. The total RFI brightness temperature in each frequency bin is computed to generate the output data cube of RFI power having dimensions $N_{altitude} \times N_{pixel} \times N_{frequency}$. Unless otherwise stated we present RFI power (strength) in Kelvin units. This data cube can be appropriately sliced to generate useful metrics of RFI(FM) in space at different locations, frequencies, and altitudes around Earth.


\noindent STARFIRE can also include global sky models such as GMOSS or improved GSM, and global CD/EoR signal models, to simulate sky spectra as would be observed in orbit around Earth, thus effectively combining beam weighted astrophysical radio-sky and FM based RFI. Such mock spectra can be used to determine the effect of flagging on FM contaminated channels, on signal detection, or conversely used to design experiments (with appropriate beams) that may be used to avoid RFI at optimal locations and orbits. 

\begin{figure*}
\begin{subfigure}{0.32\textwidth}
   \includegraphics[width=\linewidth]{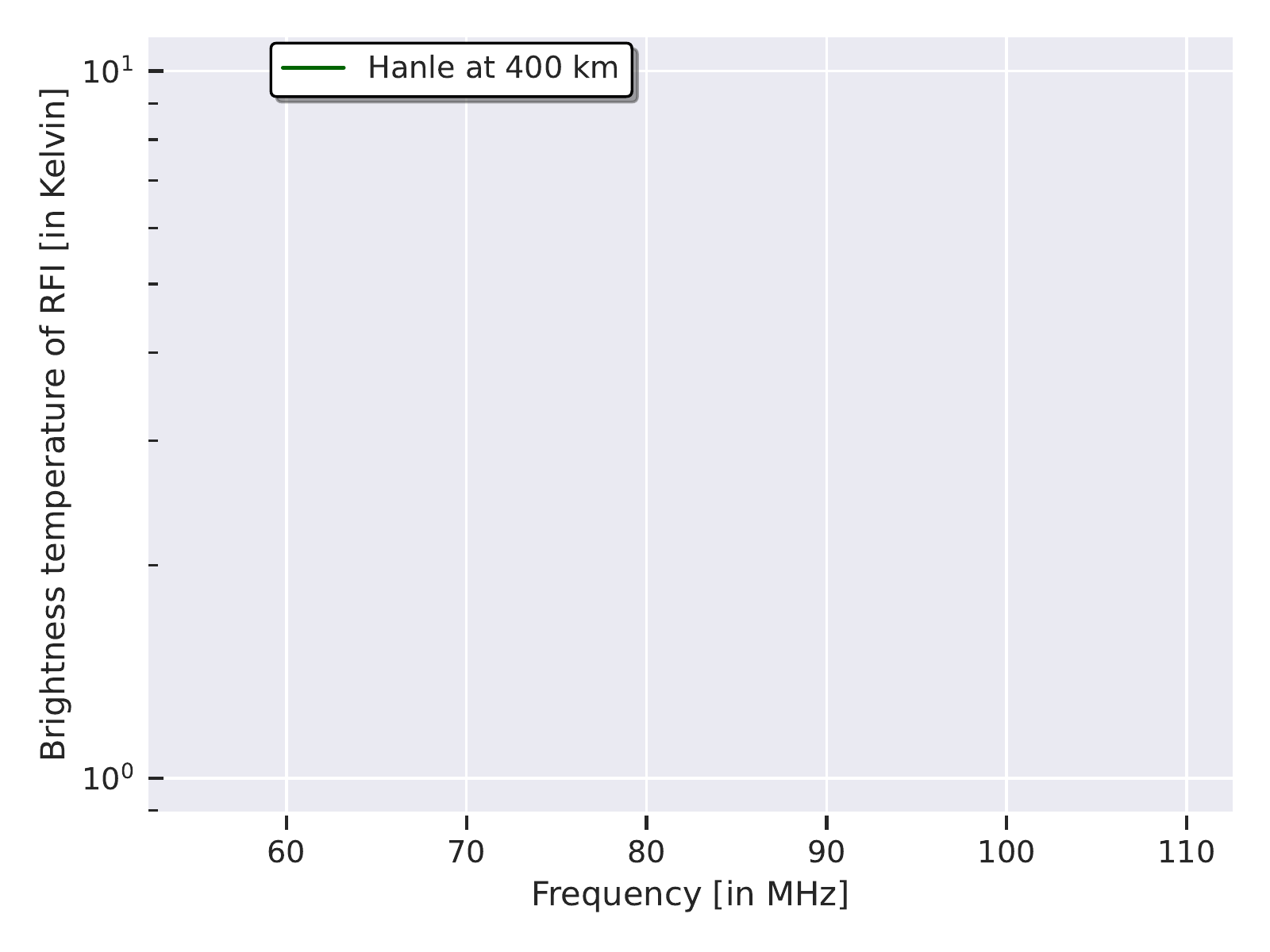}
   \caption{} \label{fig:x_a}
\end{subfigure}
\hspace*{\fill}
\begin{subfigure}{0.32\textwidth}
   \includegraphics[width=\linewidth]{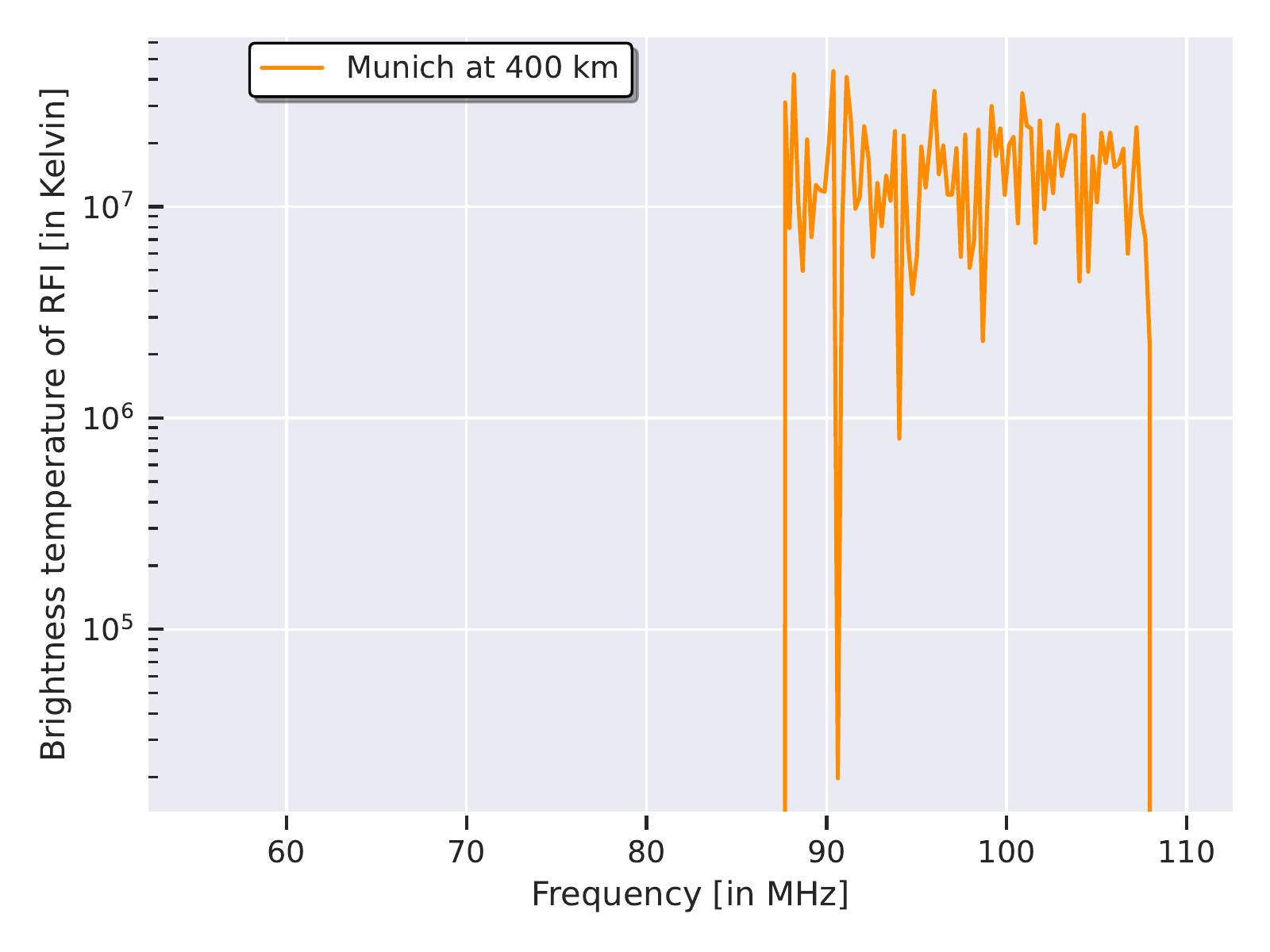}
   \caption{} \label{fig:x_b}
\end{subfigure}
\hspace*{\fill}
\begin{subfigure}{0.32\textwidth}
   \includegraphics[width=\linewidth]{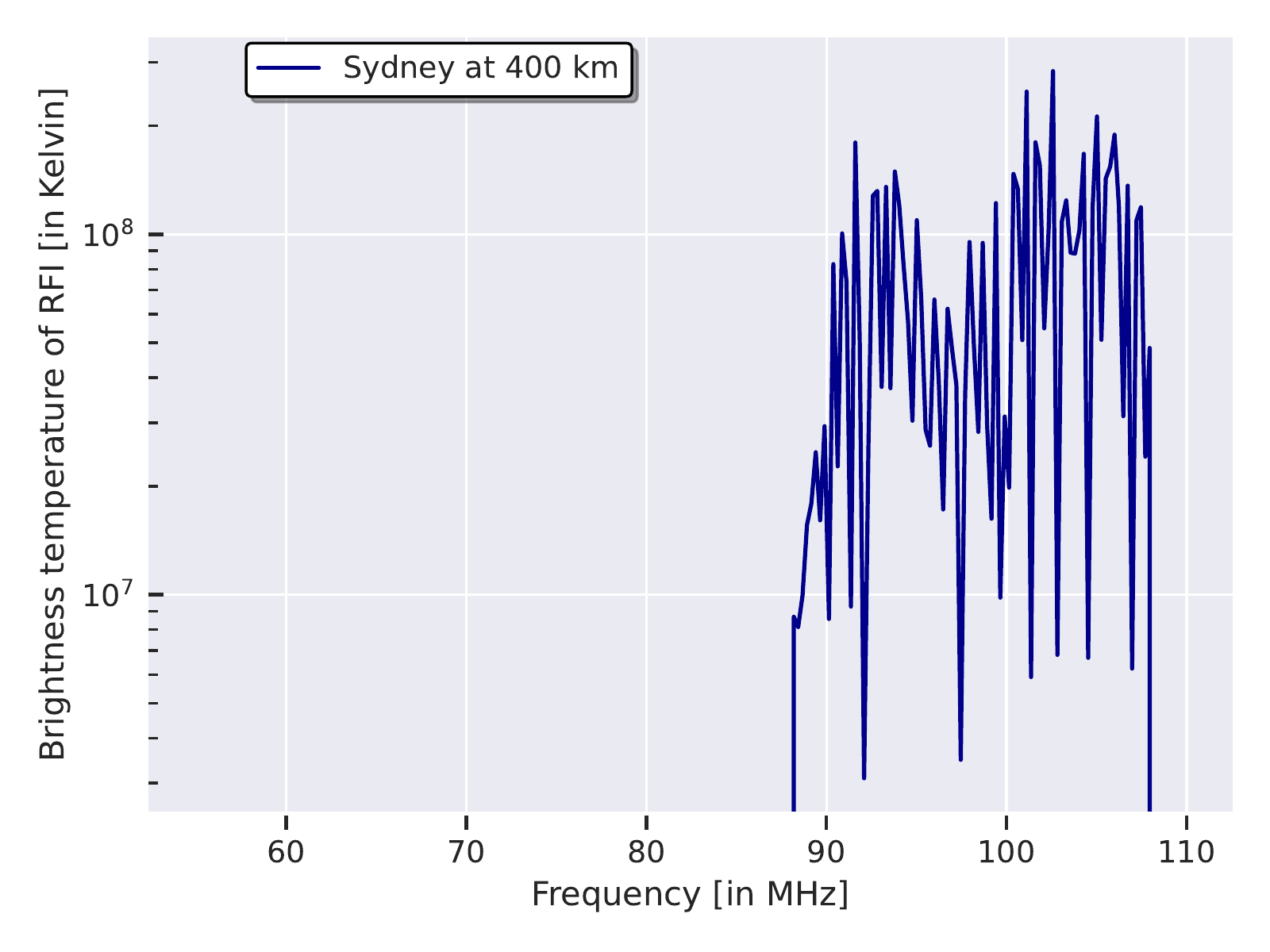}
   \caption{} \label{fig:x_c}
\end{subfigure}

\bigskip
\begin{subfigure}{0.32\textwidth}
   \includegraphics[width=\linewidth]{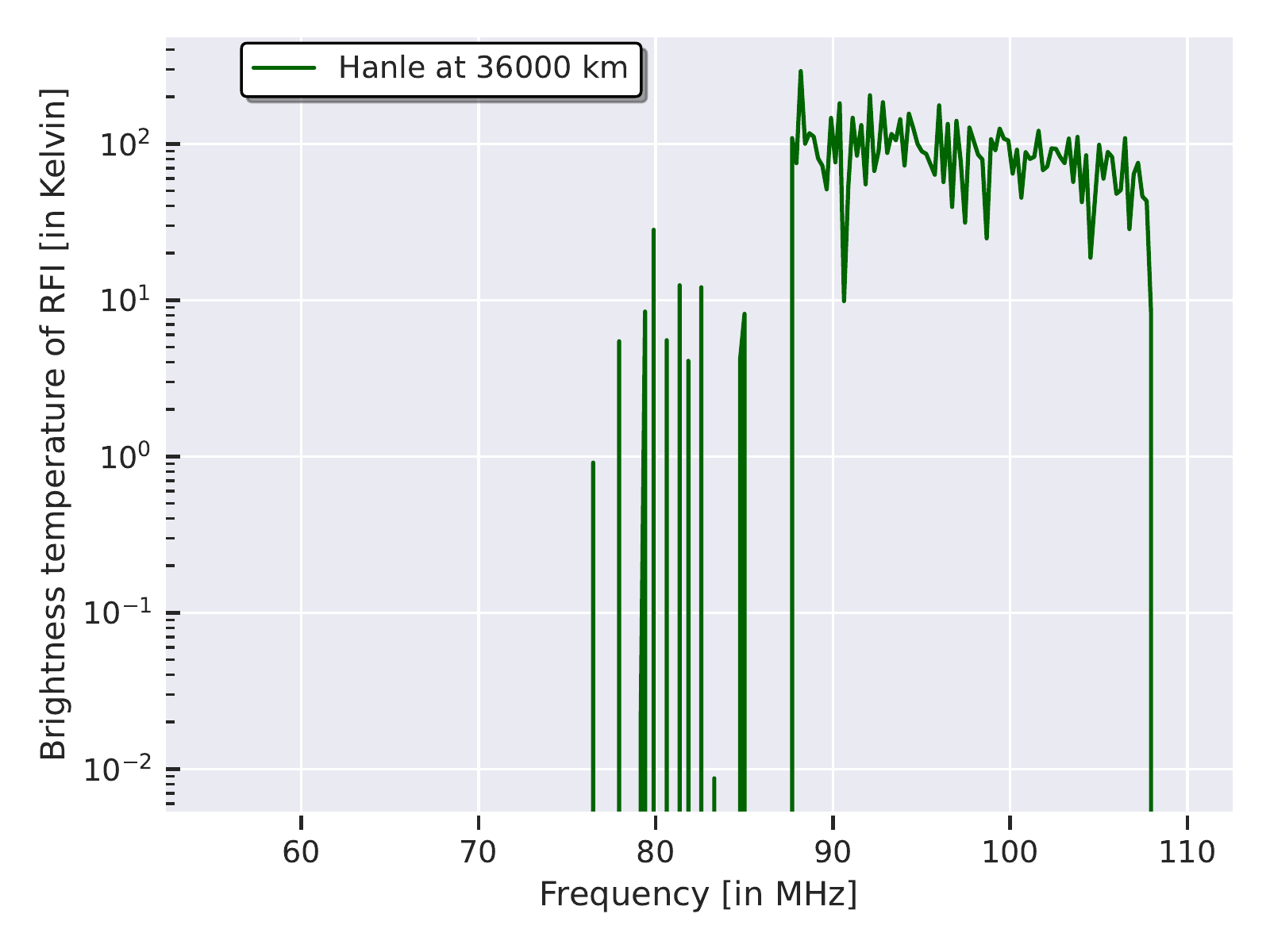}
   \caption{} \label{fig:x_d}
\end{subfigure}
\hspace*{\fill}
\begin{subfigure}{0.32\textwidth}
   \includegraphics[width=\linewidth]{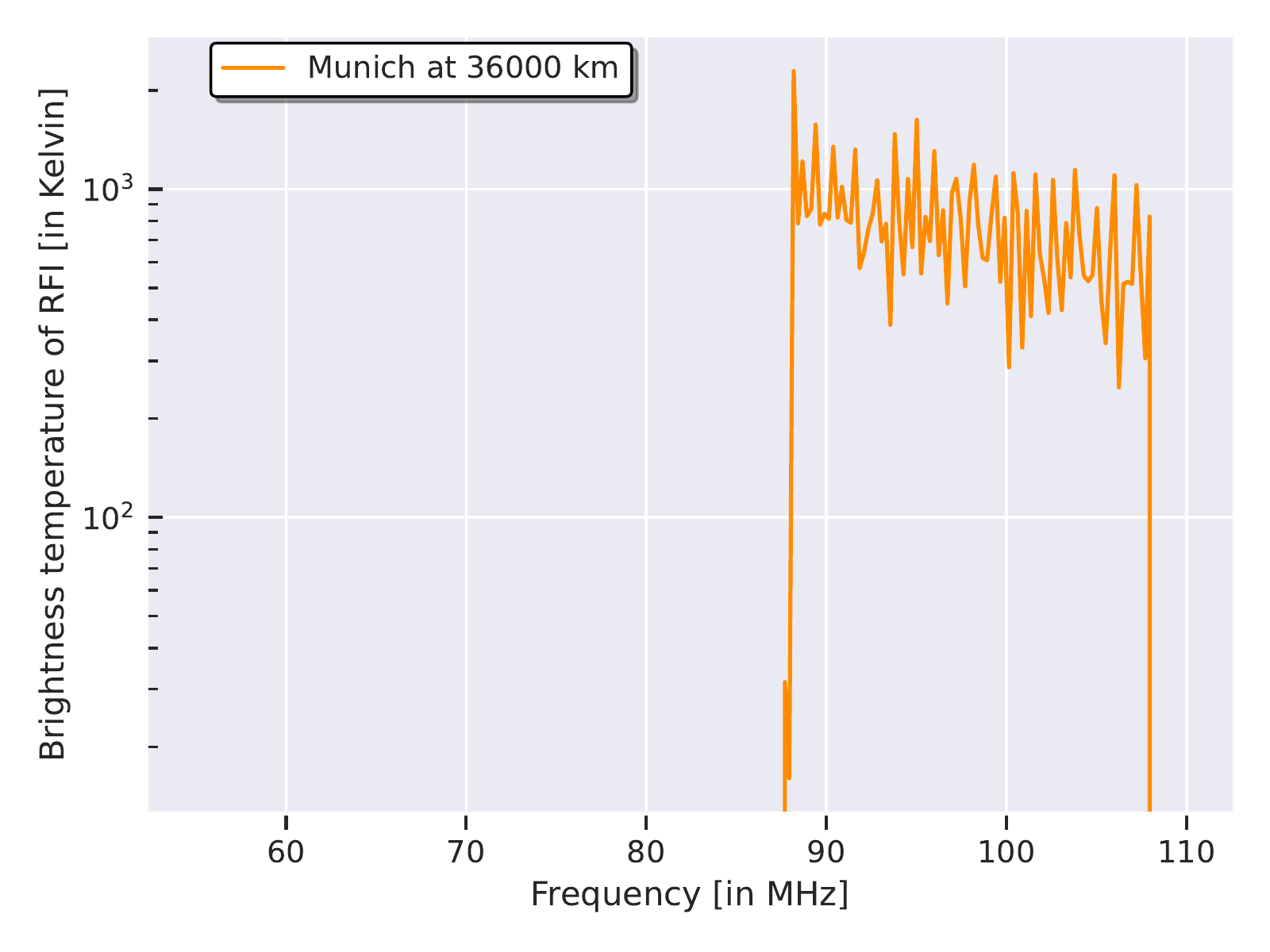}
   \caption{} \label{fig:x_e}
\end{subfigure}
\hspace*{\fill}
\begin{subfigure}{0.32\textwidth}
   \includegraphics[width=\linewidth]{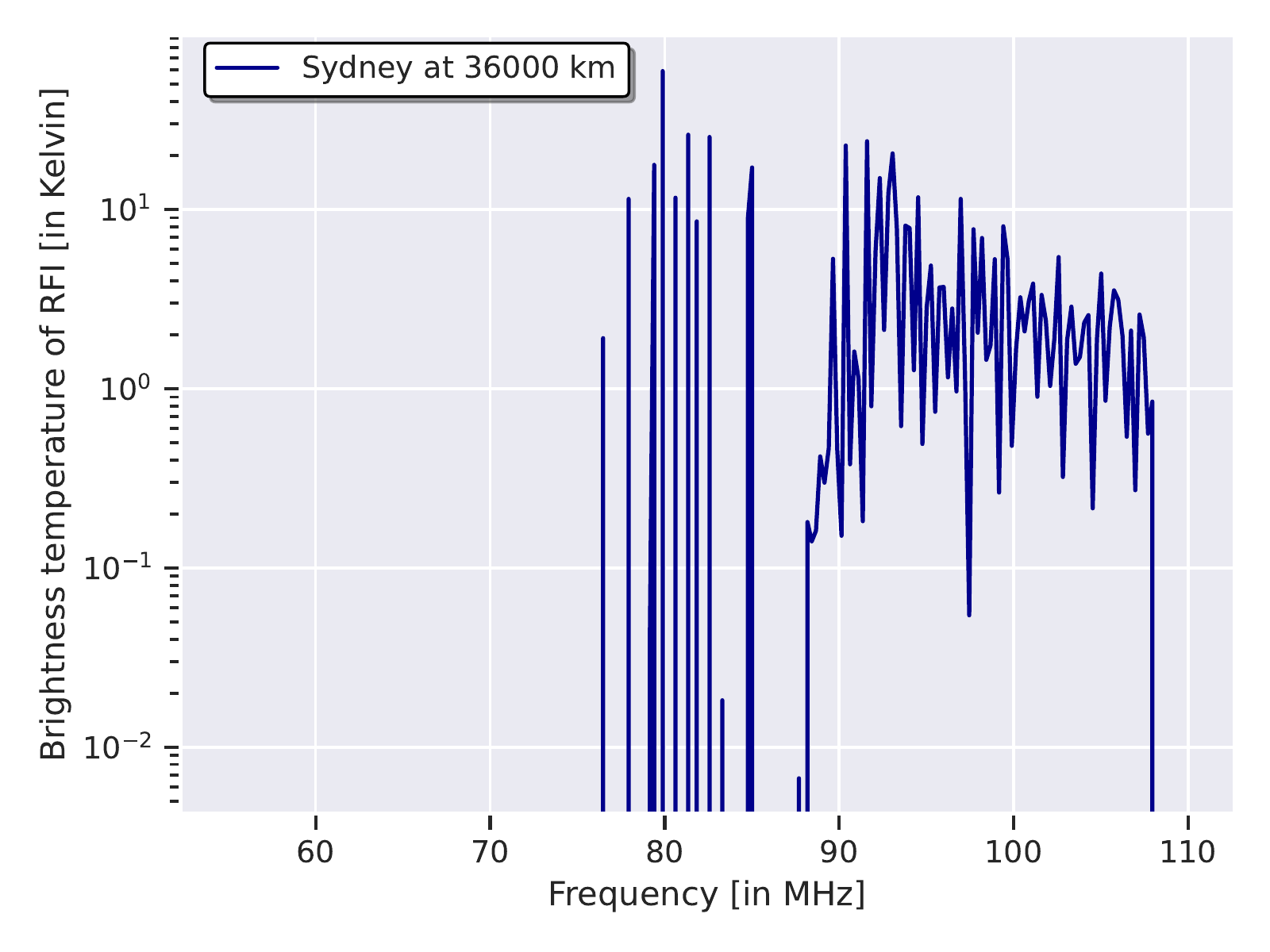}
   \caption{} \label{fig:x_f}
\end{subfigure}

\caption{Plots showing simulated RFI spectra in brightness temperature (in log scale) against frequency at 400 km (\textbf{top panel}) and 36,000 km altitude (\textbf{bottom panel}) over Hanle, Munich and Sydney from left to right respectively for a $cos^2\theta$ receiving antenna beam. In some cases RFI occupancy is observed below 88 MHz. This is due to the fact that the database contains FM transmitters from Tokyo and Australia where the FM radio band is defined between 76 to 95 MHz and 87.5 to 108 MHz respectively.}
\label{fig:spectrum}

\end{figure*}

\section{RESULTS}\label{sec:results}
\noindent STARFIRE generates a three-dimensional data cube. These can be sliced along different axes to generate different useful metrics of both RFI occupancy and RFI amplitude.
\subsection{RFI spectrum}
\noindent RFI (FM) power is expected to change based on the number of FM transmitters in the FOV, the frequency under consideration and altitude of the satellite. Figure \ref{fig:spectrum} shows the spectrum of the RFI power when the satellite is overhead different locations namely -- Hanle (India), Munich (Germany), and Sydney (Australia) -- at 400 km and 36000 km, assuming a $cos^2\theta$ beam. These spectra are obtained by slicing the three-dimensional data cube along the altitude - pixel number plane. We observe from the plots that the satellite observes significantly higher RFI power at 36000 km compared to no power at an altitude of 400 km over Hanle. This is attributed to the fact that the FM transmitter database is incomplete and does not have data for India. Thus, when the satellite is at lower altitudes, it receives no power overhead Hanle. When the satellite is at a higher altitudes it intercepts RFI from FM transmitters from almost half the Earth, encompassing the countries where the database is populated. 
This provides a consistency check for the results generated by the algorithm. For the other locations shown -- Sydney and Munich -- we note the following. RFI occupancy changes as a function of altitude, indicating contributions from different FM transmitters to different frequency bins based on the FOV. Furthermore, the total power reduces as one moves to higher altitudes as expected from the Friis transmission equation, but not perfectly so due to additional contributions from transmitters that are in the field of view for higher altitudes compared to fewer number of transmitters at lower altitudes. This demonstrates that while it is expected that the RFI power does reduce at larger distances, the channel occupancy and amplitude over different locations is a complex combination of multiple parameters and justifies the necessity of building an accurate model of the RFI environment in orbits around Earth. This quantitative estimate can help in  planning space-based radio frequency experiments orbiting Earth, including cosmology experiments targeting the global 21-cm signal from CD/EoR.

\subsection{RFI heatmap}
\begin{figure*}
\begin{subfigure}{0.32\textwidth}
   \includegraphics[width=\linewidth]{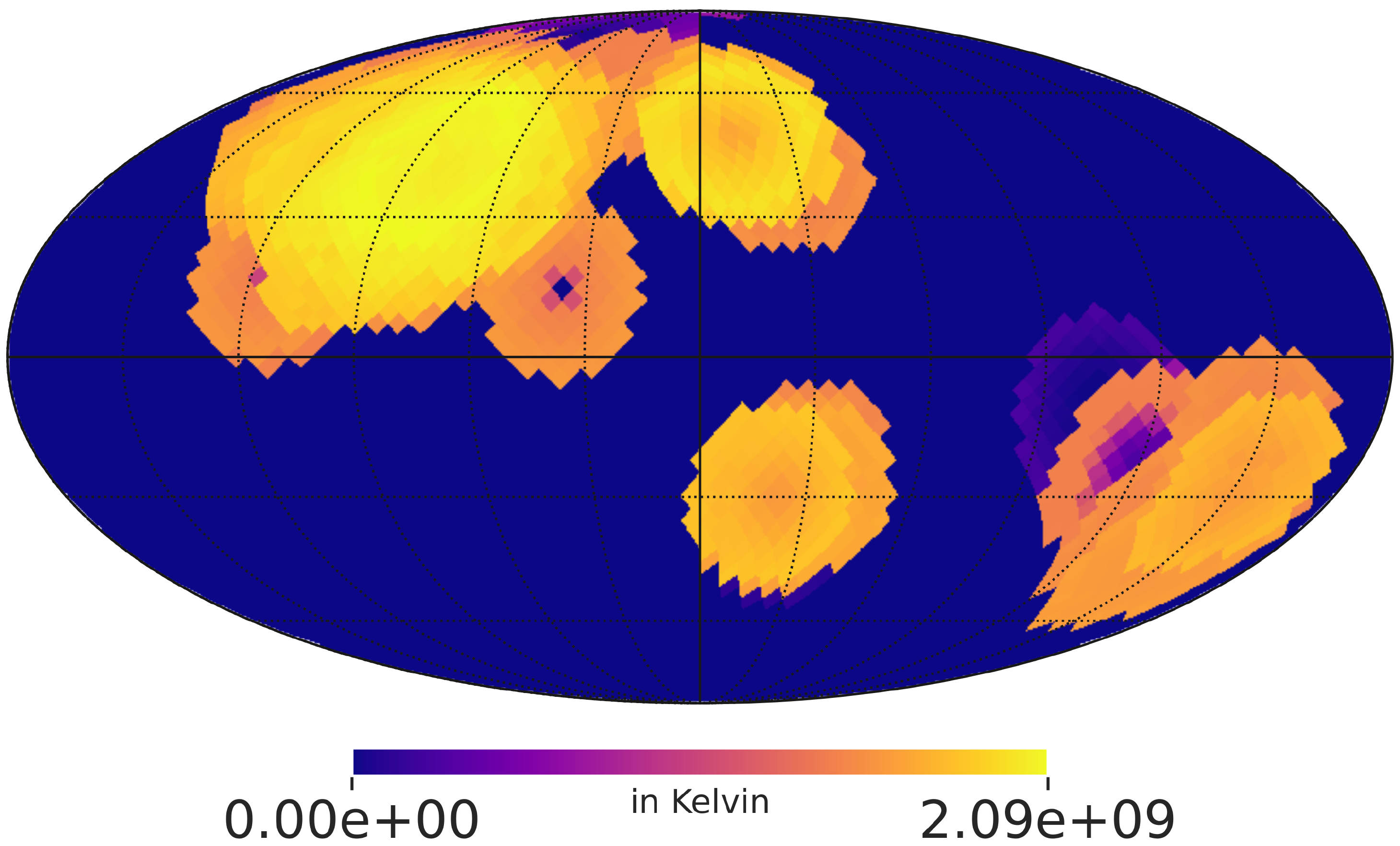}
   \caption{} \label{fig:x1_a}
\end{subfigure}
\hspace*{\fill}
\begin{subfigure}{0.32\textwidth}
   \includegraphics[width=\linewidth]{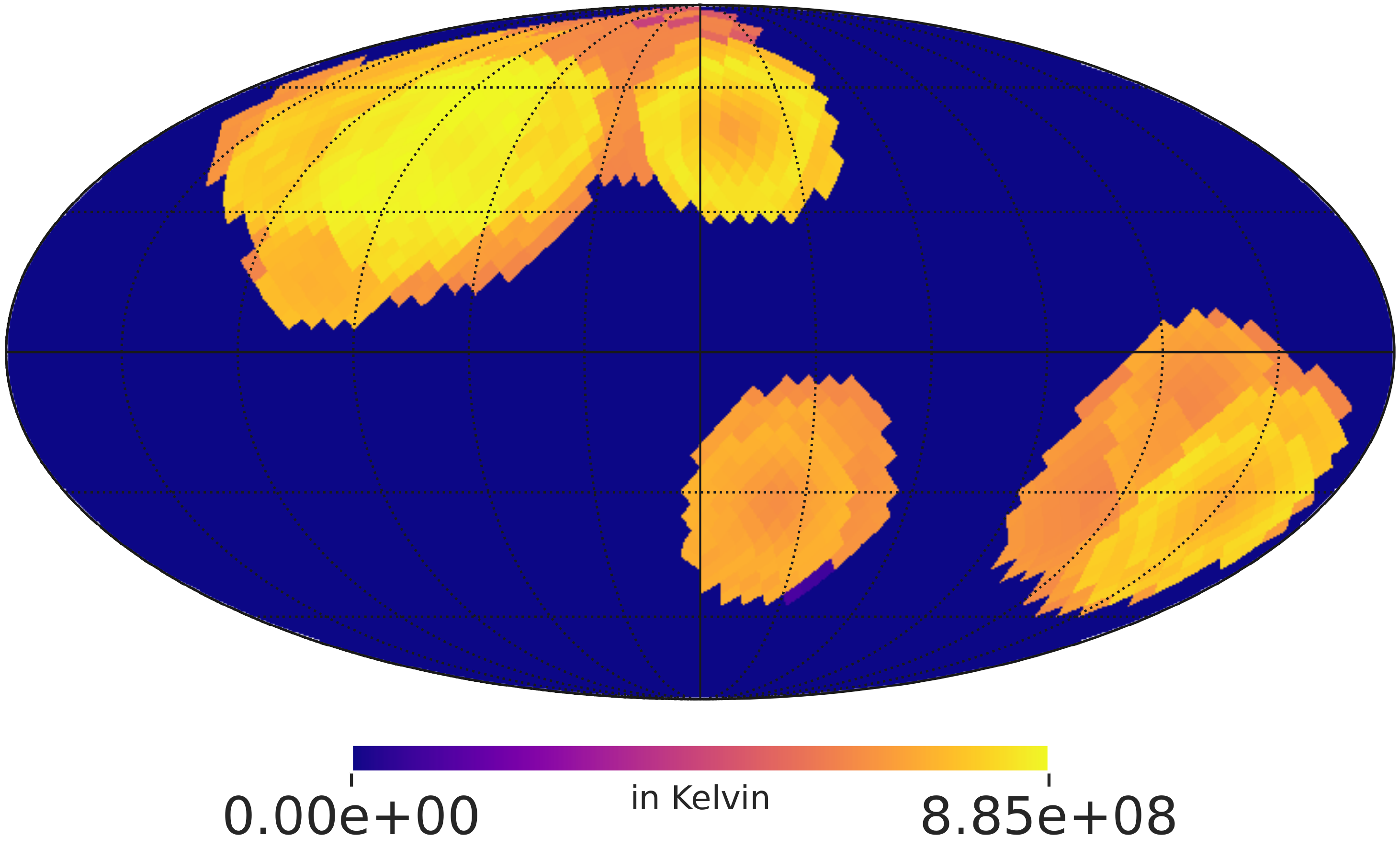}
   \caption{} \label{fig:x1_b}
\end{subfigure}
\hspace*{\fill}
\begin{subfigure}{0.32\textwidth}
   \includegraphics[width=\linewidth]{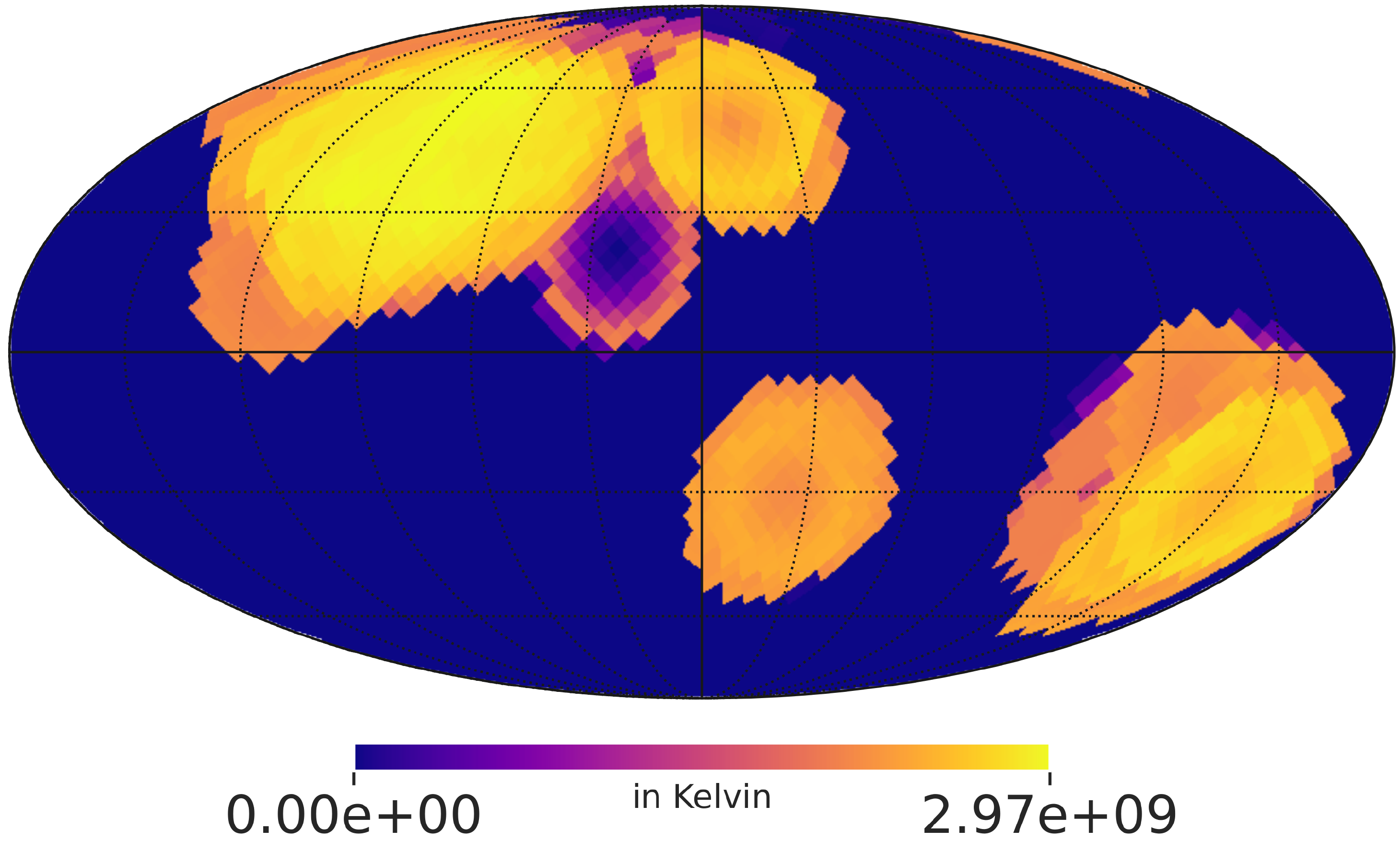}
   \caption{} \label{fig:x1_c}
\end{subfigure}

\bigskip
\begin{subfigure}{0.32\textwidth}
   \includegraphics[width=\linewidth]{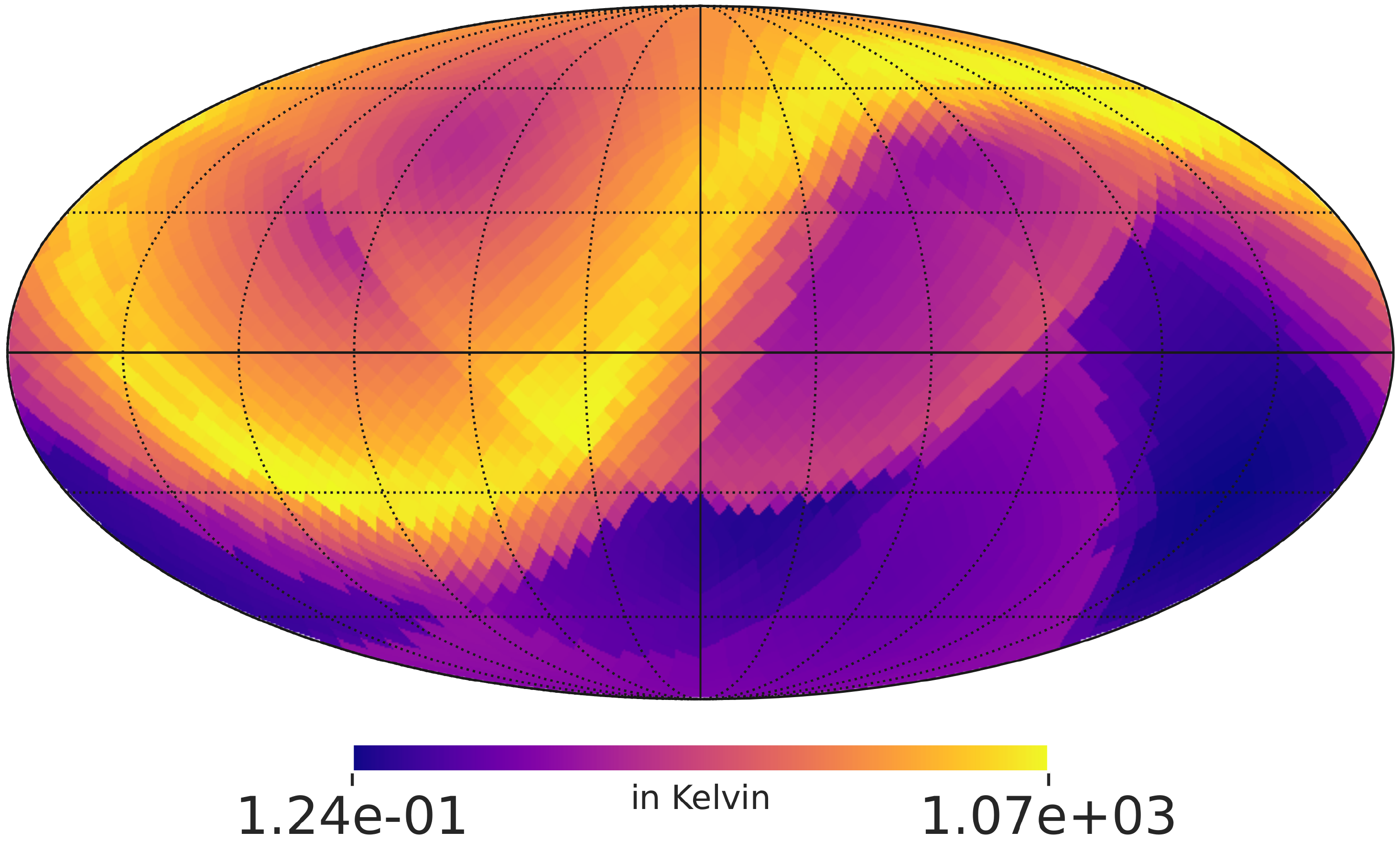}
   \caption{} \label{fig:x1_d}
\end{subfigure}
\hspace*{\fill}
\begin{subfigure}{0.32\textwidth}
   \includegraphics[width=\linewidth]{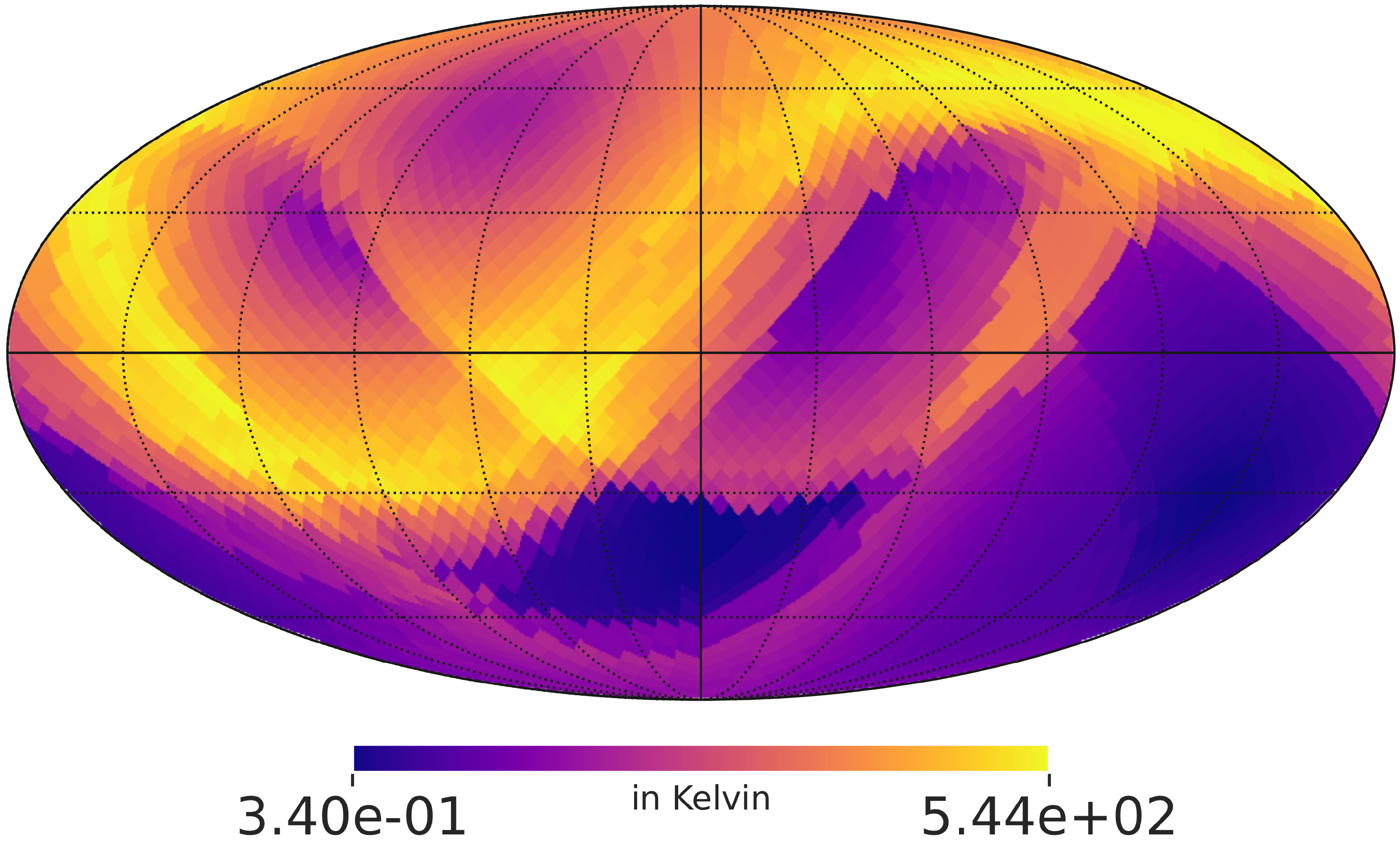}
   \caption{} \label{fig:x1_e}
\end{subfigure}
\hspace*{\fill}
\begin{subfigure}{0.32\textwidth}
   \includegraphics[width=\linewidth]{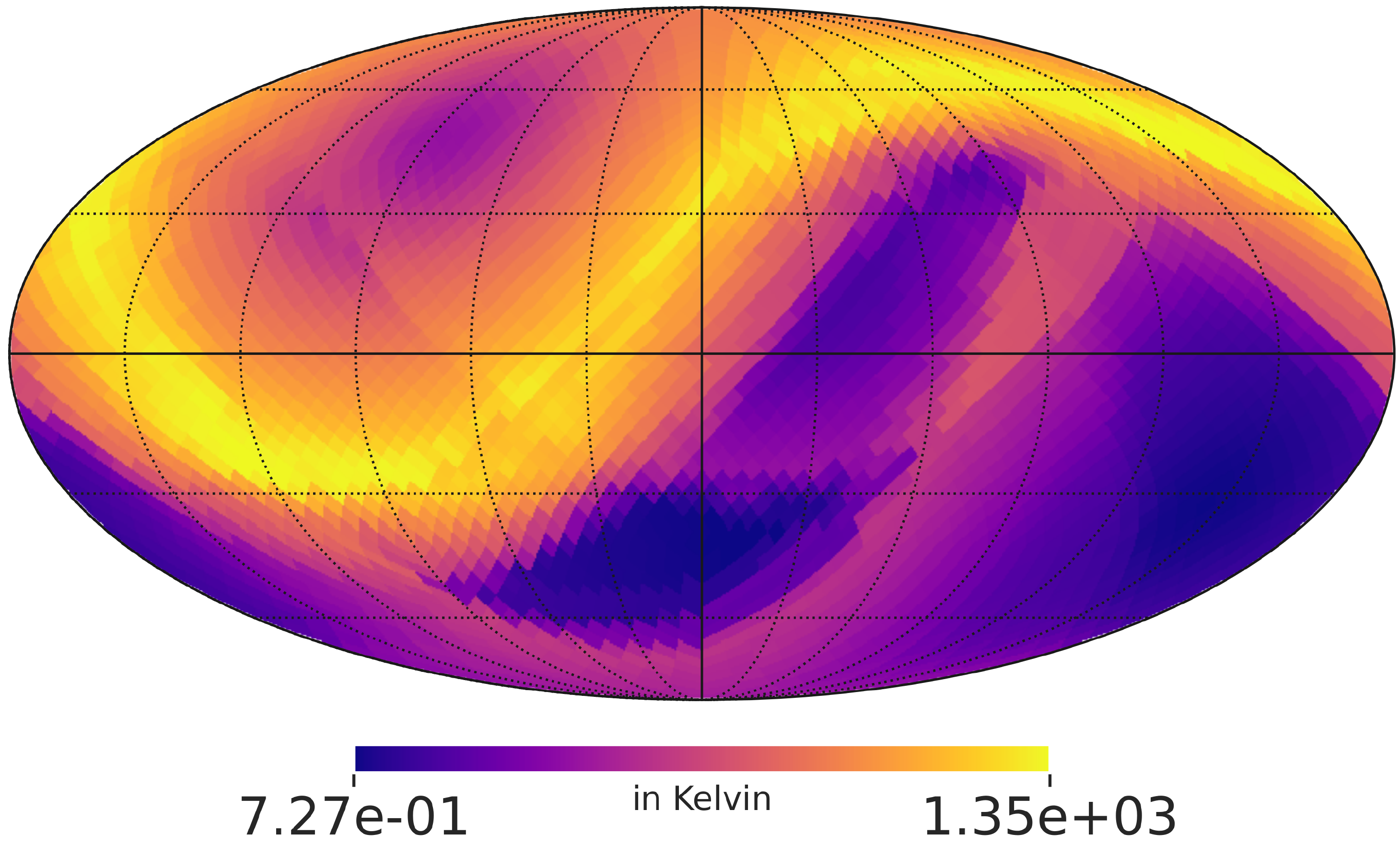}
   \caption{} \label{fig:x1_f}
\end{subfigure}

\caption{ Heatmaps for the total beam-averaged RFI power per-pixel at 400 km (\textbf{top panel}) and at 36000 km (\textbf{bottom panel}) for 90.14 MHz (left), 100.14 MHz (center) and 107.22 MHz (right). The color scale represents the strength of the RFI power (in Kelvins) received by a receiver antenna at a specific height from the FM transmitters when overhead each pixel or point on the Earth's surface, neglecting the effects of diffraction and Doppler effect. The radiation pattern of the receiving antenna and the FM transmitters were assumed to be $cos^2\theta$ and isotropic respectively. The dark blue pixels represent 0 Kelvin RFI, and correspond to those outside the beam-weighted FOV.}

\label{fig:heatmap}
\end{figure*}

\noindent Another data product that can be extracted from the three dimensional STARFIRE output data cube are heatmaps showing the total beam averaged RFI power per-pixel where pixels denote the satellite position over the globe. Such heatmaps can be generated for any of the frequency bins and altitudes over which the algorithm is run, by appropriately slicing the data cube. Figure \ref{fig:heatmap} presents the spatial variability in total RFI power (in equivalent Kelvin units) for the frequencies 90.18 MHz, 100.92 MHz and 107.75 MHz respectively at 400 km (top panel) and 36000 km (bottom panel). The code can be readily modified to generate heatmaps for any receiver antenna beam pattern as defined by the user. The "patchiness" of the heatmaps at lower altitudes is attributed to the incomplete database. As several geographical regions do not have corresponding entries in the sample database, there are large swaths of 0K RFI (indicated by the blue background) in the corresponding heatmaps, when the field of view is small at lower altitudes. At higher altitudes, the FOV increases, and a larger region is observed at the same pixel. This results in a more `filled' heatmap for the same frequency. The heatmaps in Figure \ref{fig:heatmap} are indicative of the usefulness of STARFIRE that allows for a quick inspection of the gradient of the expected RFI (FM) power over vast landscapes. A fully populated database or an equivalent realistic statistical model of FM transmitters will provide the true RFI heatmap as would be observed from space in Earth orbit. Among other applications, these heatmaps can find application in tracing the locations of the satellite with the least RFI strength at any frequency (or frequencies) and thus in planning orbits and optimizing the observation strategy -- making sky observations when passing over the low-RFI regions, for space-based CD/EoR experiments around Earth. 


\subsection{Total RFI Power vs Altitude}
\begin{figure}
    \centering
    \includegraphics[width=9.54cm]{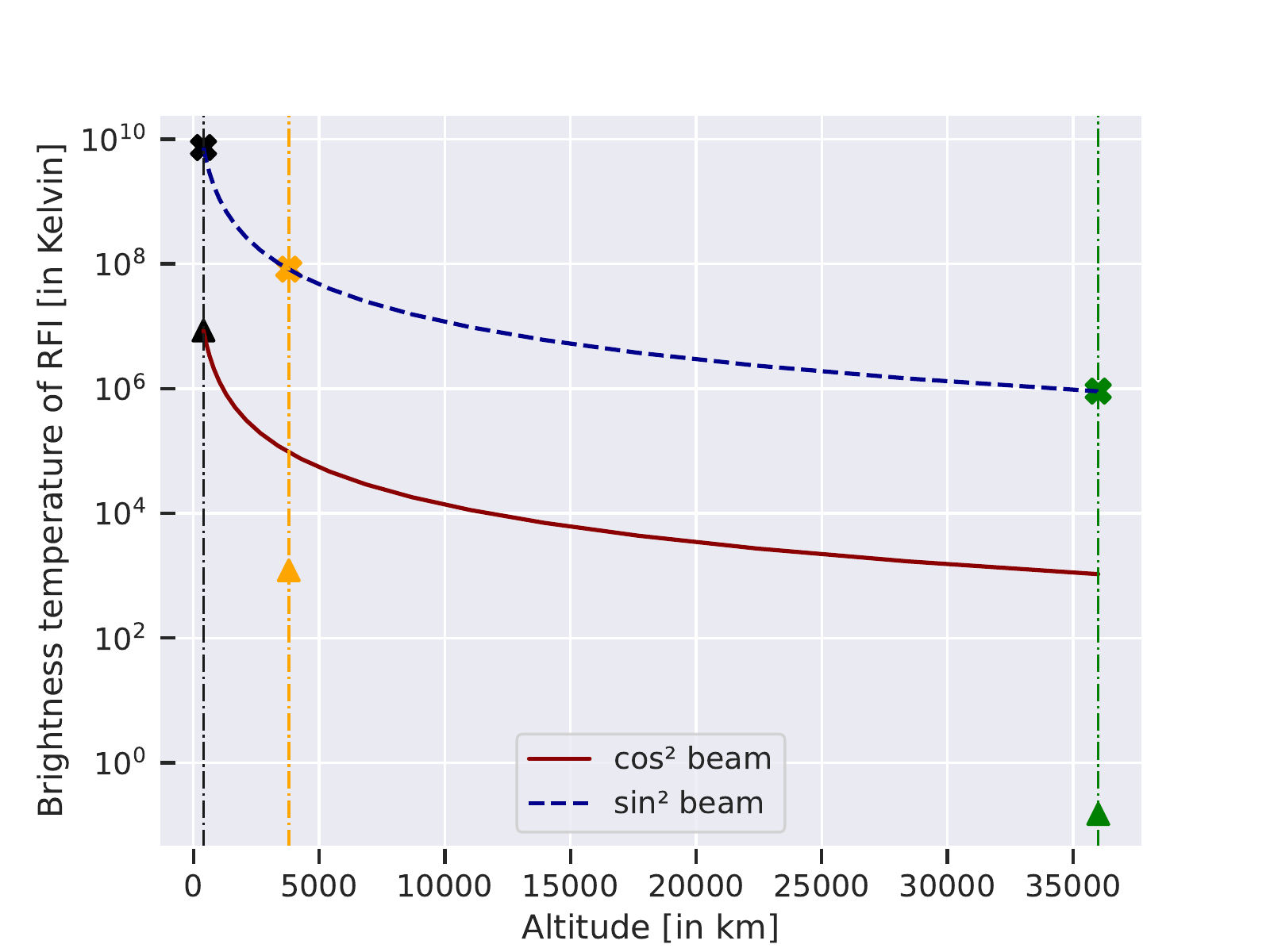}
    \caption{A scatter plot showing the RFI strength in Kelvins (in log scale) as a function of altitude at 90.14 MHz when the satellite is overhead Sydney. The data points shown in cross and triangle are the estimated powers from the actual dataset when observed by a receiving antenna having $sin^2\theta$ and $cos^2\theta$ beam patterns respectively. The solid and dashed lines correspond to the expected RFI levels when following the Friis transmission equation (Equation \ref{eq:FT}) and extrapolating the power at 400 km to higher altitudes, for the two different beams. It is noted that for the $sin^2\theta$ beam pattern, the RFI power estimated at an altitude of 36000 km is $\sim$ 7.5$\times 10^{3} $ Kelvin above the theoretical estimate from the Friis equation (refer Table \ref{tab:beam_comparison_table}). The calculated RFI strength for $cos^2\theta$ beam pattern is observed to be $\sim$ 1.0$\times 10^{3} $ Kelvin below the Friis-extrapolated power at the same altitude.}
    \label{fig:pow_alt}
\end{figure}

\noindent Another useful result of STARFIRE is the total RFI strength as a function of altitude over any given location on Earth. This can in turn be the total power across the full frequency band or at a specific frequency bin. At first glance it might appear that RFI power reduces as a function of altitude and this is indeed true. However, as the FOV determines the number of transmitters whose powers as intercepted by the receiver are summed varies with altitude, the total power levels do not strictly follow the Friis transmission formula. This is demonstrated in Figure \ref{fig:pow_alt}. The total  RFI strength overhead Sydney at 90.18 MHz is estimated for two different receiving antenna beam patterns, namely $sin^2\theta$ and $cos^2\theta$. The plot demonstrates that when we have an Earthward looking beam pattern, in this case the $sin^2\theta$ beam pattern with a maximum directed towards Sydney, the RFI power received by the satellite at higher altitudes is 8.2$\times 10^{6}$ Kelvin more than what is expected from the scaling of the Friis equation at 400 km. This is because the satellite has more number of transmitters in its FOV even though there is considerable amount of free-space attenuation with increasing altitude. On the other hand, with a $cos^2\theta$ beam pattern with a null towards the subsatellite point, there is an additional reduction of RFI power (represented by the triangular data points) than expected for the same altitudues from the interpolation of the Friis equation. Although the satellite has larger FOV owing to the altitude, the effective number of transmitters in the beam-weighted FOV is small as the peak of the receiver beam pattern is directed mostly along the azimuthal plane direction with lesser number of transmitters.

\section{Discussion}\label{sec:discussion}
\subsection{Figure of Merit}\label{sec:FOM}
\noindent For the specific application of STARFIRE to PRATUSH we define a FOM at each location (altitude, latitude, longitude) as the effective RFI free bandwidth. To the first order, our definition of FOM emphasizes the RFI channel occupancy, wherein the RFI contaminated channels are flagged and hence reduce the effective bandwidth. Specific to PRATUSH, a 4-term Nutall window function is employed in the correlation spectrometer, such that extremely strong RFI (the threshold of which is set by the number of bits in the digital receiver) can result in spillover of the otherwise localized RFI to the entire band, resulting in the loss of the full spectrum. In this case too, our definition of FOM applies, wherein the effective bandwidth is zero. Thus, when using STARFIRE to identify optimal orbital parameters for an Earth Orbiting CD/EoR experiment, the orbit that maximizes cumulative FOM over the total mission lifetime will be most favorable. 

\begin{figure*}
\begin{subfigure}{0.53\textwidth}
   \includegraphics[width=\linewidth]{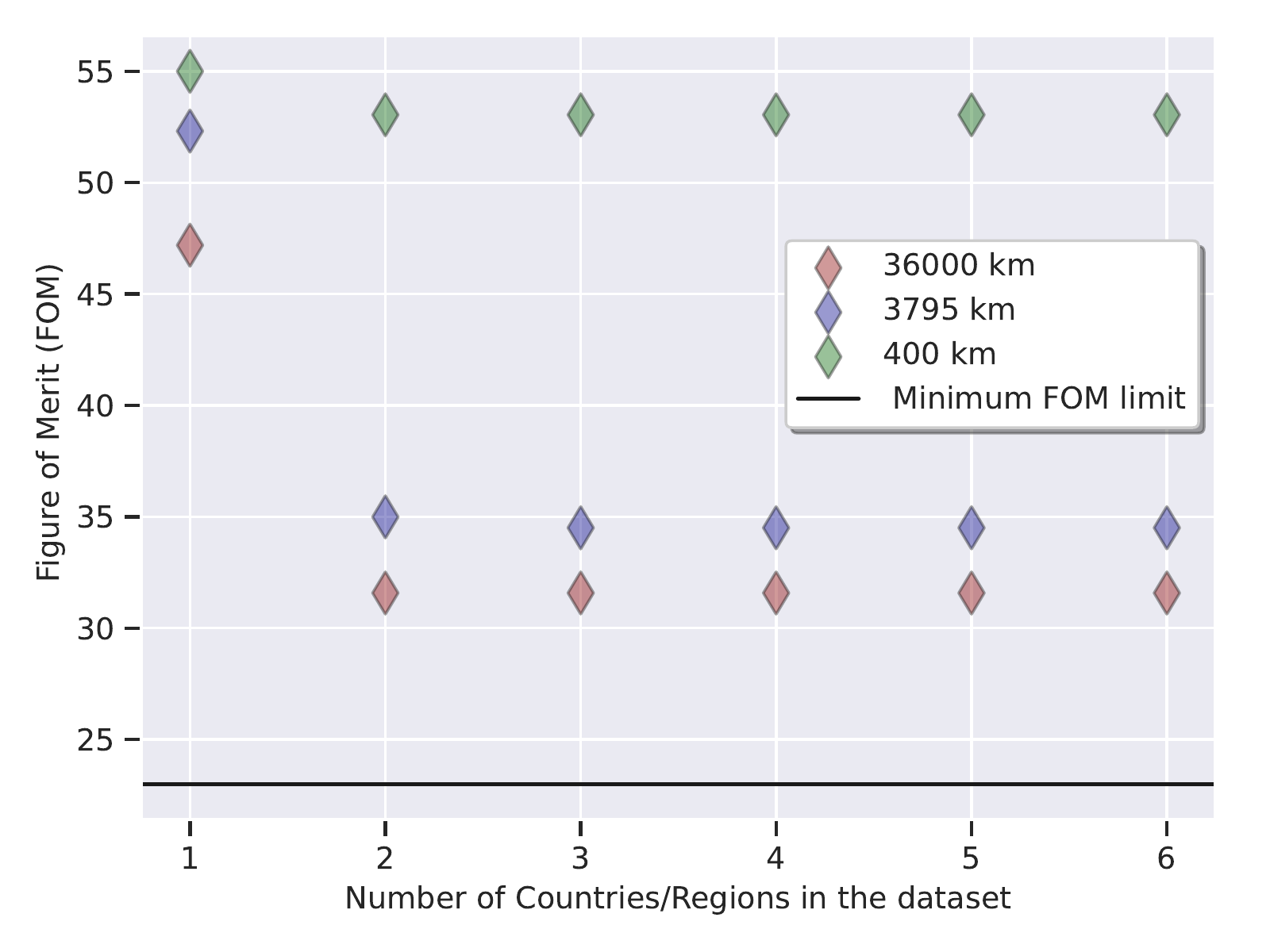}
   \caption{Location over Arctic Ocean (83.785306\degree N, 45.494139\degree E)} \label{fig:xxx_c}
\end{subfigure}
\begin{subfigure}{0.53\textwidth}
   \includegraphics[width=\linewidth]{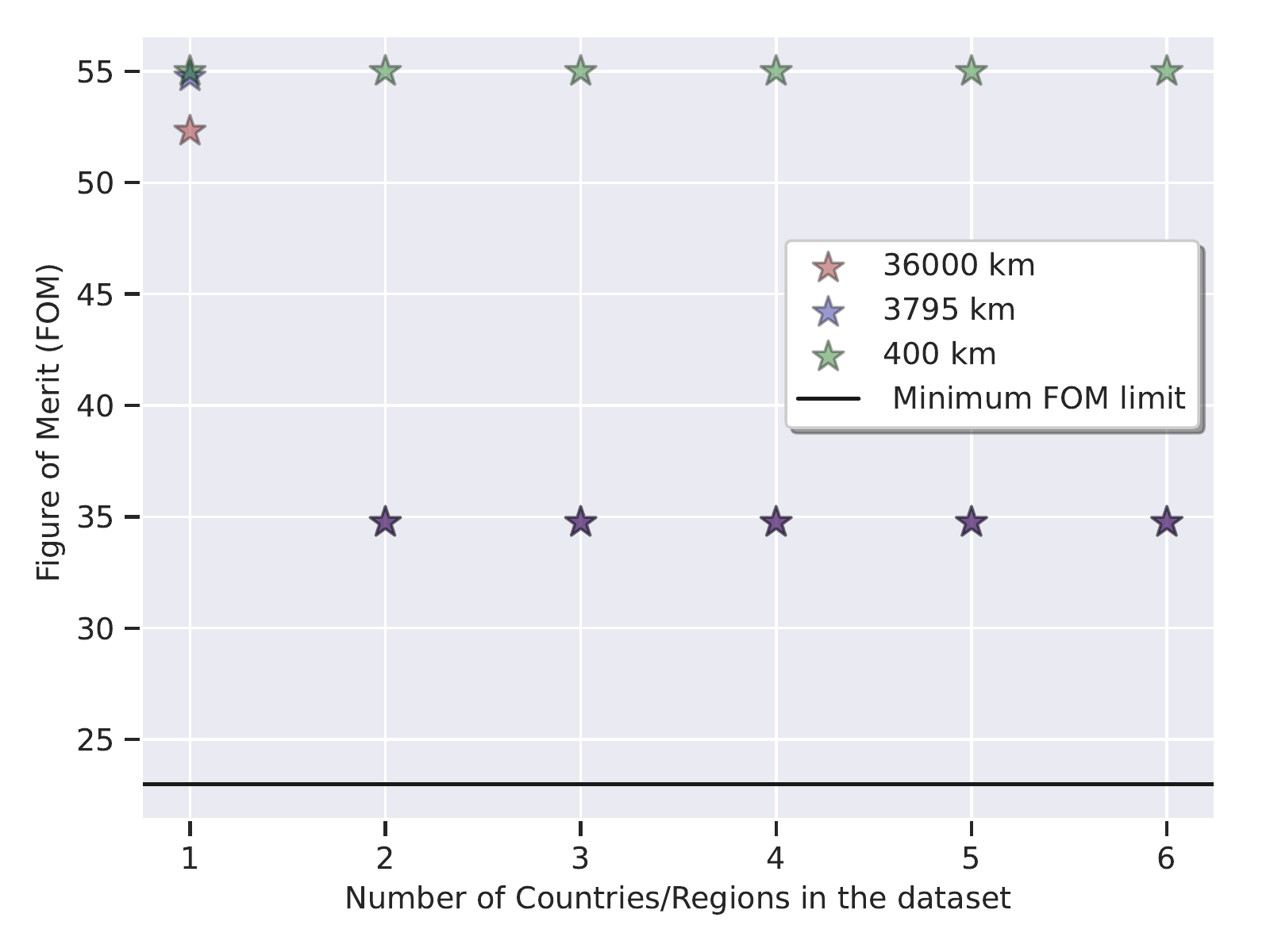}
   \caption{Location over Pacific Ocean (8.7832\degree S, 124.5085\degree W)} \label{fig:xxx_d}
\end{subfigure}

\caption{FOM at different locations at 400 km, 3795 km and 36000 km when varying the number of regions in the limited FM transmitter database.}
\label{fig:FOM}
\end{figure*}

With this definition of FOM, and with the limited database currently available and assuming RFI (FM) as the sole determinant of orbital parameters, one would deduce that a low altitude (400 km) orbit maximizing time over oceans would satisfy the criteria of maximizing FOM.
Figure \ref{fig:FOM} shows the variation of FOM when the satellite is over Arctic Ocean (83.785306\degree N, 45.494139\degree E) and Pacific Ocean (8.7832\degree S, 124.5085\degree W). To estimate the consequence of the limited database on the output of STARFIRE, we see the variation in the FOM deduced as transmitters from different regions (countries) are introduced one at a time, at each of the three default altitudes. Understandably the FOM deteriorates towards higher altitudes, with more transmitters in the FOV and presumably more complete spectral coverage from the database, with the steepest drop in the FOM when countries or regions with the most complete records are included. Assuming the worst case scenario of the FM band fully occupied over 76-108 MHz (note that while traditionally the FM band occupies 88-108 MHz, the FM transmission in Japan extends down to 76 MHz) the theoretical minimum FOM for PRATUSH is 23. Despite the limited number of transmitters in our presently limited database, due to their geographic distribution across the globe, at an altitude of 36,000 km when half the Earth is in the field of view, the fractional error between the theoretical limit and the calculated FOM is 37.29\% and 51.08\% over the Arctic Ocean and and over the Pacific Ocean pixels respectively. For more detailed calculations, please refer to~\ref{app}. These large errors result from the incomplete database, such as data missing altogether from South America. For instance, data missing from South America is reflected in the blue patches in the heatmaps even at the large altitude of 36,000 km, as shown in the bottom panel of Figure 
 \ref{fig:heatmap}.



We also note here that FOM for other applications of STARFIRE can be suitably determined by the users.

\subsection{Expected RFI levels}
\noindent With the caveat that decades old measurements of RFI from space will not hold true today due to the increase in the usage of the spectrum for communication and FM transmitters, we undertake a brief exercise in comparing these measurements with estimates of RFI levels computed by STARFIRE to obtain a back-of-the-envelope confidence check of the order of magnitude levels against the existing literature.

Measurements from Radio Astronomy Explorer 1 (RAE-1)\citep{weber1971radio} at $9.18$ MHz showed that observed RFI levels were $\sim 44$ dB above the galactic background at an altitude of $6000$ km. We proceed to determine an order of magnitude estimate of the expected RFI level at $88$ MHz, that defines the lower bound of the assumed FM frequency band. Assuming the brightness temperature of the radio sky to be $248$ K at $150$ MHz \citep{landecker1970galactic} and a spectral index of $\alpha = -2.5$ \citep{bowman2018absorption} we derive the radio sky brightness to be 940.75 K at 88 MHz. We assume that the relative strength of RFI compared to the galactic emission determined by RAE-1 in orbit at 6000 km is also valid at other altitudes and follows the Friis transmission equation. At 400 km, the RFI level at 88 MHz is $10log_{10}(\frac{6000}{400})^2$ times that at 6000 km, since it would be observed closer to the source of RFI (Earth surface). The the estimated RFI level in equivalent brightness temperature would be $\sim 5.32\times10^{9}$ K. On the other hand at 36000 km the expected RFI strength in equivalent temperature units is estimated to be $\sim 6.56\times10^{5}$ K.

RAE-2 was placed in lunar orbit operating in the frequency range of 250 kHz to 13.1 MHz. The data from RAE-2 show that the radiation level from terrestrial RFI sources were about 10 dB to 20 dB above the galactic noise, the range emerging from RFI variations on Earth between day and night. Assuming a mean Earth and Moon distance of 384000~km, we estimate that at 400 km, the RFI level would be increased by $10log_{10}(\frac{384000}{400})^2$= 59.64 dB resulting in a maximum RFI equivalent temperature of $\sim 8.67\times10^{10}$ K. For an altitude of 36000 km, the power will be $10log_{10}(\frac{384000}{36000})^2$= 20.56 dB times that in the lunar orbit resulting in an equivalent brightness temperature of the RFI to be $\sim 1.07\times10^{7}$ K.

Comparing these back of the envelope estimates with the values of the data points (from the real dataset) in Table \ref{tab:beam_comparison_table}, we see that the estimated values are similar in order of magnitude to the values of the RFI power for a satellite having a receiver antenna with $sin^2\theta$ radiation pattern. This provide further confidence in the order of magnitude RFI brightness estimated with STARFIRE.

\begin{table}
	\centering
	\caption{The calculated (from actual dataset) and predicted RFI power overhead Sydney (in orders of Kelvin) for different beam patterns and altitudes at a frequency of 90.14 MHz}
	\label{tab:beam_comparison_table}
	\begin{tabular}{l|c|c} 
		\hline
		Beam Pattern & Actual Data & Friis formula \\
		\hline
		$cos^2\theta$ at 400 km & 8562341.41 K & 8562340.00 K \\
		$sin^2\theta$ at 400 km & 7320965643.92 K & 7320965640.00 K \\
		$cos^2\theta$ at 36000 km & 0.15 K & 1057.08 K \\
		$sin^2\theta$ at 36000 km & 911411.00 K & 903822.92 K \\
		
		\hline
	\end{tabular}
\end{table}

		

\subsection{Effect of RFI on radio sky spectrum}
\begin{figure*}

\begin{subfigure}{0.50\textwidth}
   \includegraphics[width=\linewidth]{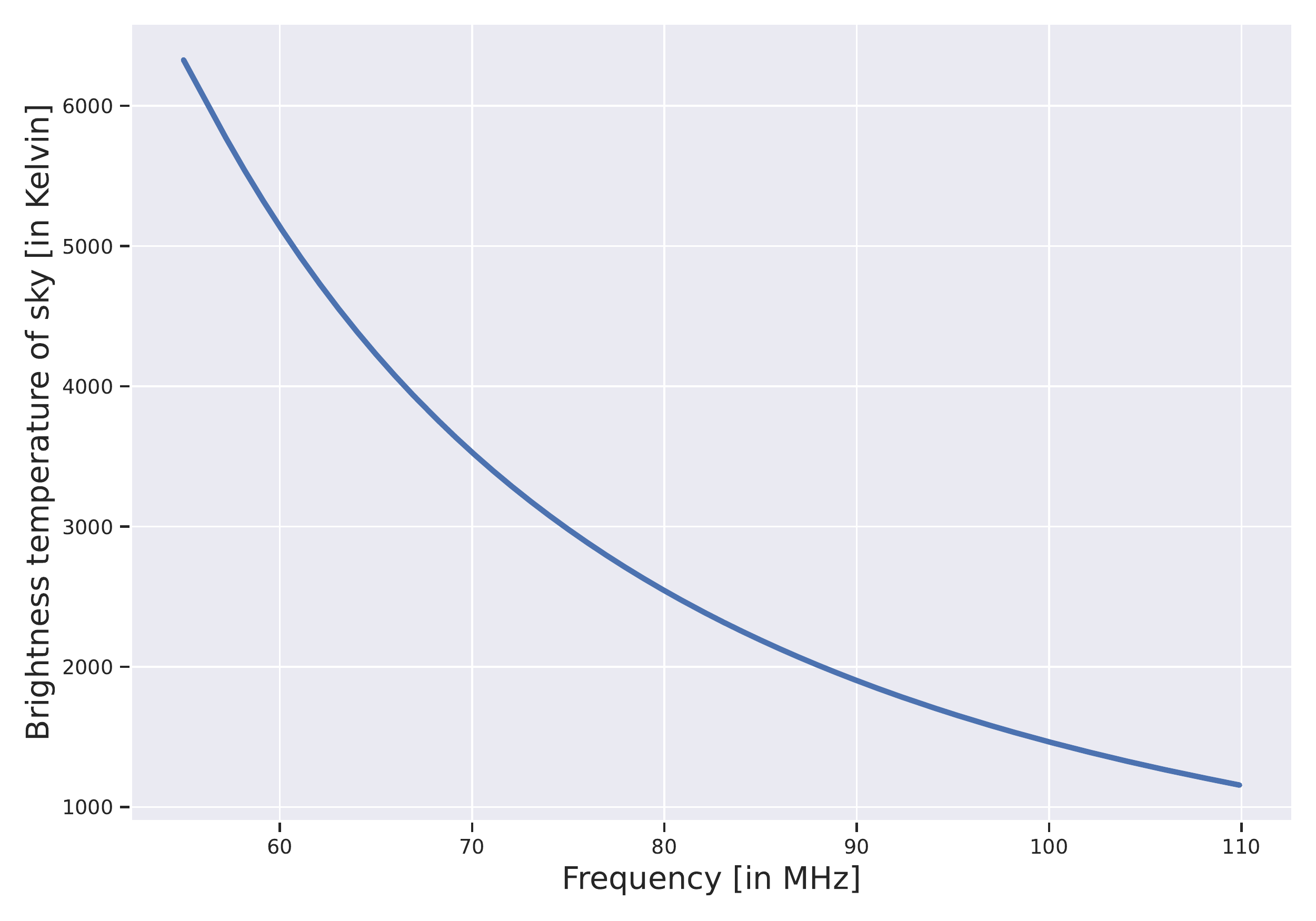}
   \caption{} \label{fig:xx_c}
\end{subfigure}
\begin{subfigure}{0.50\textwidth}
   \includegraphics[width=\linewidth]{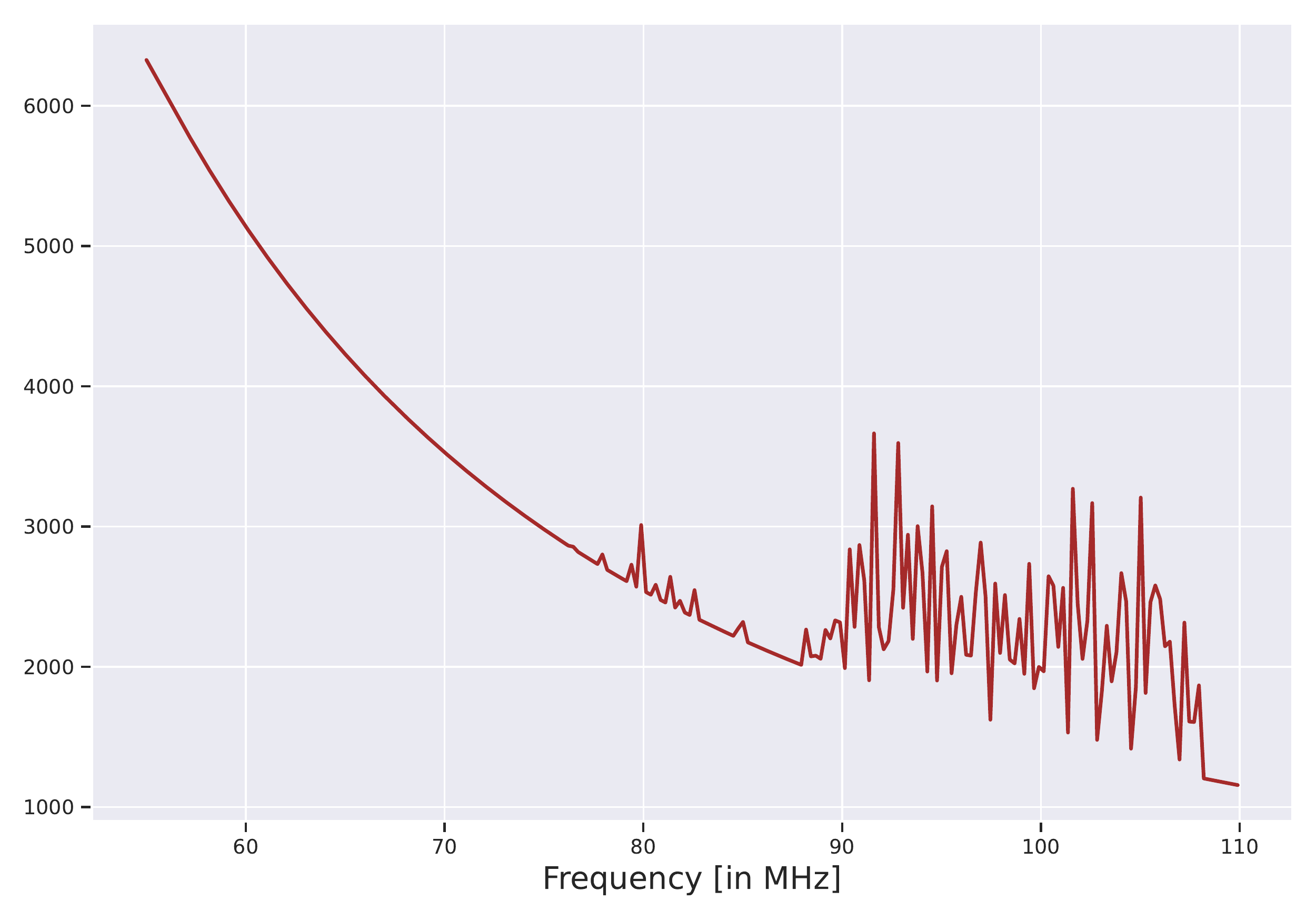}
   \caption{} \label{fig:xx_d}
\end{subfigure}

\caption{ Figure \ref{fig:xx_c} displays the mock astrophysical sky spectrum over Sydney at an altitude of 36000 km as observed by the PRATUSH antenna and adopting GMOSS as the radio sky model. When RFI estimated from STARFIRE is included the observed sky spectrum changes to that shown in Figure \ref{fig:xx_d}. }
\label{fig:sky_rfi}
\end{figure*}

In the context of detecting the redshifted global 21-cm signal, STARFIRE can be readily adopted to include a radio sky model in addition to the RFI model. For instance, STARFIRE can include global sky models such as GMOSS or improved GSM, and global CD/EoR signal models. This can be used to simulate sky spectra as would be observed by an experiment with user specified antenna properties, in orbit around Earth (see Figure \ref{fig:sky_rfi}), effectively combining beam-weighted radio-sky containing the cosmological signal with RFI from the FM radio band. Such mock spectra can be used to determine the required level of flagging in FM contaminated channels at different locations and altitudes. This can be used to determine total channel loss and consequently the effect of flagging on signal detection. This analysis will also facilitate the determination of the most suitable orbits for Earth orbiting global CD/EoR detection experiments and conversely be used to guide experiment design by way of optimization of antenna beams with a maximized skyward gain away from the terrestrial RFI. 

To demonstrate generating mock spectra as would be observed by a CD/EoR detection experiment we use the PRATUSH baseline configuration as an example. Adopting the notation in Figure \ref{fig:el}, we convolve the beam skyward (from the antenna defined horizon) with the GMOSS radio-sky. We then convolve the antenna beam Earthward (beneath the antenna defined horizon) with the RFI model as described previously. Figure \ref{fig:sky_rfi} shows mock sky spectra thus generated over Sydney at 36000 km without any RFI (for instance as would be observed in an ideal no RFI scenario) and the effect of including RFI in the same spectrum for the same LST. Clearly, several channels are affected by RFI and would be flagged in an experiment seeking to detect the redshifted global 21-cm signal from CD/EoR. The flagging procedures, and the effect of channel loss on CD/EoR signal detection confidence is beyond the scope of this paper and will be presented in a follow up paper.


\section{Conclusions and Future Work}\label{sec:summary}
STARFIRE is an algorithm that can be used to quantitatively estimate the strength of RFI in space over Earth. Using a geographically distributed, limited database of FM transmitters, we have presented a pipeline that can generate a three dimensional data cube as its output. This output can be sliced appropriately to generate useful metrics of RFI occupancy. There are several default parameters that STARFIRE assumes, such as frequency range, frequency bins, altitudes, antenna beams for both receiving and transmitting antennas, to name a few. These can be readily modified by the user. STARFIRE can find applications in several areas, where knowledge of RFI in space is required. Of particular interest in our case is the effect of RFI on experiments seeking a detection of the redshifted global 21-cm signal from CD/EoR in Earth orbits. For such an application, STARFIRE can be used to identify optimal orbits with low RFI occupancy around Earth and aid space-based experiment design. 

One might intuit that the orbits that spend the most amount of time at low altitude over oceans with presumably the least number of transmitters in the field of view might maximize FOM. However, a space-based experiment operating in such an orbit might see significantly larger RFI occupancy compared to a ground-based experiment that has an advantage of limiting RFI using line-of-sight propagation. We foresee that for the highest confidence CD/EoR signal detection, the optimal method is to operate two complementary experiments, one operating in space and the other on the ground. While the ground based experiment has the advantage of relatively low RFI, the Earth orbiting experiment would be devoid of chromatic effects induced by the ionosphere and electromagnetic coupling of antenna with the dielectric medium. A signal seen in data sets of both experiments would make a strong case for high-confidence detection! This the strategy that would be employed by PRATUSH phase I.

There are several areas of development which are the focus of STARFIRE-2. This includes providing a statistical model for FM transmitters across the globe in lieu of a comprehensive database from each country. For the case of global-21 cm detection experiments, a more complete database can be used to determine suitable low-RFI orbits with appropriate ground trace, or inform the customization of the antenna beam to minimize the total terrestrial RFI as would be viewed from orbit. Furthermore, statistical modeling can further aid in improving the spatial resolution of the database using appropriate mathematical interpolation techniques. Whereas the STARFIRE framework is flexible to include any source of RFI that can be described in the same format as the sample database of FM transmitters herein, the current STARFIRE framework does not provide a direct formalism to account for the time (and hence position) variable source of RFI such as satellites. Examining the effect of RFI contaminated spectra with spatial, temporal and frequency variations on signal detection and cosmological parameter estimation will be presented in a follow up paper.
\section{Acknowledgements}\label{sec:acknowledge}
S.G. would like to thanks Jishnu N.T, Mugundhan Vijayaraghavan, and Vishwapriya Gautam for their valuable inputs.

\section{Data availability}\label{sec:data}
The data cube generated for the results presented in this paper, along with STARFIRE, will be made accessible to the public at \texttt{https://github.com/distortions21cm/STARFIRE}. \\
The database sample used in the current implementation of STARFIRE are available from sources in the public domain as listed below.
\begin{enumerate}
  \item Australia: \url{https://www.acma.gov.au/list-transmitters-licence-broadcast}
  \item Canada: \url{https://www.ic.gc.ca/engineering/BC_DBF_FILES/baserad.zip}
  \item Germany: \url{https://en.wikipedia.org/wiki/List_of_FM_radio_stations_in_Germany}
  \item South Africa: \url{https://www.icasa.org.za/publication-of-final-terrestrial-broadcasting-frequency-plan-2008}
  \item Japan(Tokyo): \url{https://fccdata.org/?zone=JP&latd=35.66340&lond=139.69576&lang=en}
  \item USA: \url{https://www.fcc.gov/media/radio/fm-query}

\end{enumerate}

\appendix
\section{The estimated error in the FOM }
\label{app}
\noindent We assume the minimum or worst case figure of merit (FOM) closer to the true value, since adding more data will result in the estimated FOM approaching closer to the minimum FOM. From this we estimate percent error introduced due to incompleteness of the data using the formula : 
\begin{equation}    
{\rm \%\ error} = 100 \times \frac{({\rm Estimated\ value} - {\rm True\ or\ worst\ case\ value})}{{\rm True\ or\ worst\ case\ value}}
\end{equation}

We select pixels that are geographically remote/isolated such as over the Arctic and Pacific oceans. The percentage error shown is determined using the worst FOM estimated at 36,000 km with all the six countries included, as given by the aforementioned formula and this works out to be 37.29\% and 51.08\% respectively.

Over Arctic Ocean: $(31.576-23)/23\times100\% = 37.29\%$

Over Pacific Ocean: $(34.748-23)/23\times100\% = 51.08\%$

While certainly large, the error will reduce as the database is populated.

\section{Impact of adding more transmitters}
\label{compare_data}
\begin{figure*}
    \centering
    \includegraphics[width=\textwidth]{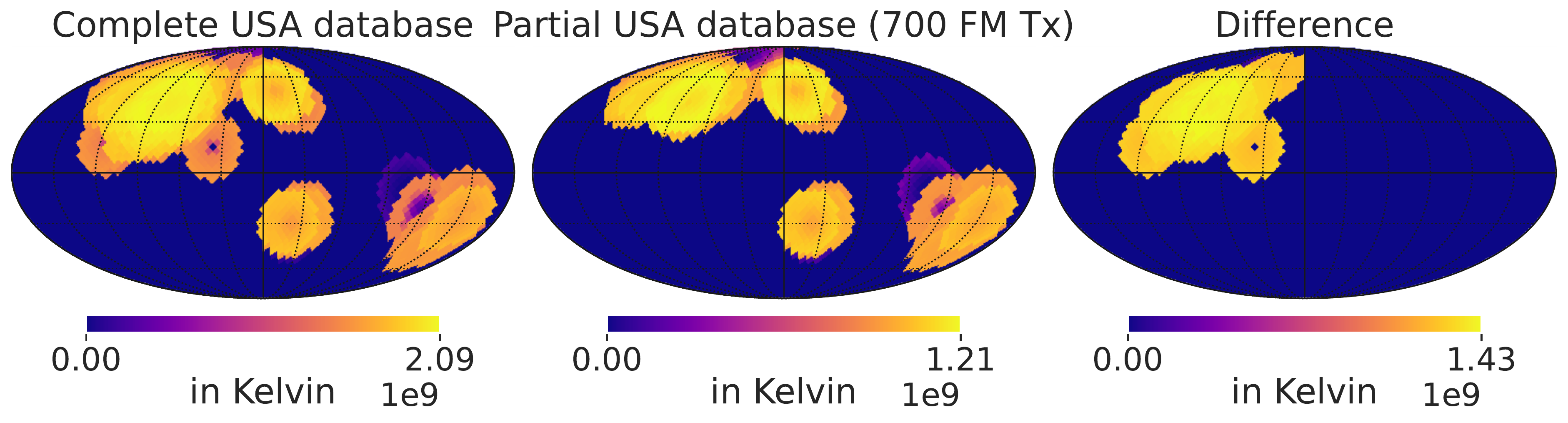}
    \caption{The figure presents heatmaps illustrating the total RFI power per-pixel at 400 km, specifically at a frequency of 90.14 MHz. The two cases being considered here involves two different FM transmitter databases. The first case employs the entire US FM transmitter dataset (\textbf{left}) in the input database, which comprises approximately 28,300 entries. In contrast, the second case utilizes a limited US database consisting of only 700 FM transmitters (\textbf{center}). We can clearly observe that the heatmaps will exhibit a higher degree of coverage as the database is populated by incorporating more number of transmitters into it. The difference (\textbf{right}) shows the extra power that gets added with the complete database.}
    \label{fig:old_new_compare}
\end{figure*}

We undertake an exercise here to show the impact of changing the number of transmitters within a country, such as the US. For instance, if the database for a country is incomplete the effect of the omission is dependent on the distribution of the transmitters. The worst case scenario is when all the transmitters considered are concentrated in one region, say only along the coasts (when in reality there are significant numbers inland) or along borders. We consider a mock scenario where we populate only 700 out of the entire US FM transmitter database ($\sim{28300}$ entries), with the former concentrated along the northern latitudes of the country. The heatmaps for the total beam-averaged RFI power per-pixel at a distance of 400 km and a frequency of 90.14 MHz are presented in Figure \ref{fig:old_new_compare}. The heatmaps clearly indicate that 

\begin{figure}[H]
    \centering
    \includegraphics[width=\linewidth]{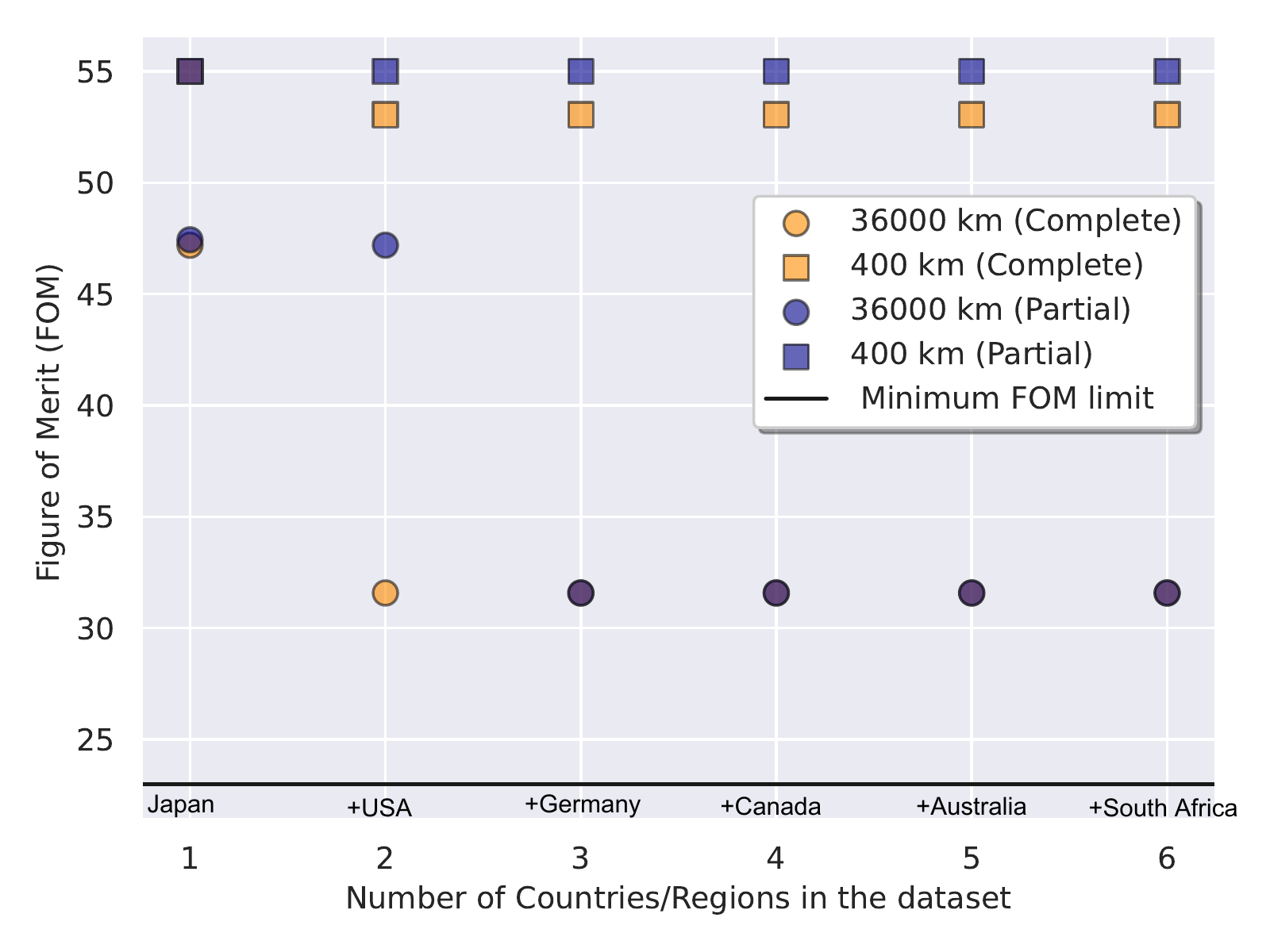}
    \caption{FOM over Arctic Ocean (83.785306\degree N, 45.494139\degree E) at 400 km and 36000 km. The theoretical minimum FOM for PRATUSH is 23. A drastic decrease in FOM decrease is observed as the database is populated with the complete USA transmitter database (in yellow) compared to a subset of 700 (in blue) .} 
    \label{fig:old_new_compare_FOM}
\end{figure}

\noindent as the FM transmitter database is expanded with input data, the  filling of heatmaps is expected at all altitudes, resulting in fewer regions of distinct blue color.

Figure \ref{fig:old_new_compare_FOM} shows the FOM over a pixel over the Arctic Ocean  (83.785306\degree N, 45.494139\degree E) at two different altitudes. Adding the complete database of USA results in a drastic degradation of FOM compared to the case when only 700 transmitters are considered. This is representative of the kinds of biases and errors that can result in a partial or incomplete database, and thus motivates building up a complete database especially for conclusive results at low altitudes.

\bibliographystyle{elsarticle-harv} 
\bibliography{bibliography}

\end{document}